\documentclass[aps,prd,twocolumn,superscriptaddress,floatfix,nofootinbib,amsmath,amssymb
]{revtex4-2}

\usepackage{aasmacros}
\usepackage[utf8]{inputenc}
\usepackage{graphicx}
\usepackage{dcolumn}
\usepackage{bm}
\usepackage{newtxtext,newtxmath}
\usepackage{lipsum}
\usepackage{subcaption}
\usepackage[T1]{fontenc}
\usepackage{makecell}
\usepackage{animate}
\usepackage{mathtools}
\usepackage{url}
\usepackage[hidelinks,colorlinks]{hyperref}
\usepackage{xcolor}
\hypersetup{
    colorlinks=true,
    linkcolor={blue},
    citecolor={blue},
    urlcolor={blue}
}

\begin{document}

\preprint{APS/123-QED}

\title{The Sensitivity of Substructure Lensing to SIDM Core-collapse Model Variation}

\author{Charlie Mace}
\email{E-mail: mace.103@osu.edu}
\affiliation{Department of Physics, The Ohio State University, 191 W. Woodruff Ave., Columbus OH 43210, USA}
\affiliation{Center for Cosmology and Astroparticle Physics, The Ohio State University, 191 W. Woodruff Ave., Columbus OH 43210, USA}

\author{Birendra Dhanasingham}
\affiliation{Minnesota Institute for Astrophysics, University of Minnesota, 116 Church St. SE, Minneapolis, MN 55455, USA}

\author{Zhichao Carton Zeng}
\affiliation{Department of Physics and Astronomy, Mitchell Institute for Fundamental Physics and Astronomy, Texas A\&M University, College Station, Texas 77843, USA}

\author{Francis-Yan Cyr-Racine}
\affiliation{Department of Physics and Astronomy, University of New Mexico, 210 Yale Blvd NE, Albuquerque, NM 87106, USA }

\author{Xiaolong Du}
\affiliation{Department of Physics and Astronomy, University of California, Los Angeles, California 90095, USA}

\author{Annika H. G. Peter}
\affiliation{Department of Physics, The Ohio State University, 191 W. Woodruff Ave., Columbus OH 43210, USA}
\affiliation{Center for Cosmology and Astroparticle Physics, The Ohio State University, 191 W. Woodruff Ave., Columbus OH 43210, USA}
\affiliation{Department of Astronomy, The Ohio State University, 140 W. 18th Ave., Columbus OH 43210, USA}
\affiliation{School of Natural Sciences, Institute for Advanced Study, 1 Einstein Drive, Princeton, NJ 08540}

\author{Andrew Benson}
\affiliation{Carnegie Science, Observatories, 813 Santa Barbara Street, California 91101, USA}

\date{\today}

\begin{abstract}
Strong gravitational lensing has emerged as a powerful probe of dark matter substructure, and shows particularly strong promise as a test of self-interacting dark matter (SIDM). The compact halos produced by SIDM can leave distinct imprints on lensing observations, but the core-collapse timeline for subhalos is difficult to model accurately. This difficulty is an obstacle to accurate substructure lensing predictions, where small variations in core-collapsing subhalos can lead to significant differences in the lensing power. To quantify this problem and inform future lensing analyses, we test various methods of modeling core-collapsing halos and show the effect of each variation on the two-point correlation function of the effective deflection field's divergence and curl. Our tests include smoothly evolving density profiles versus instantaneously collapsing halos, probabilistic collapse versus individual halo evolution, and variation of the initial and final density profile parameters. We find that the two-point correlation function is sensitive to most of these variations at small length scales, but the detectability of these differences will depend on the observational probe.

\end{abstract}

\maketitle

\section{Introduction}

Self-interacting dark matter (SIDM) is a set of particle physics models that include interactions between particles other than the gravitational force, in contrast to commonly invoked cold dark matter (CDM) models, where there is no such interaction. In SIDM models with elastic collisions, which are the focus of this study, the heat transfer facilitated by dark matter interactions will cause the characteristics of dark matter halos to evolve over time, starting with the formation of isothermal, constant-density cores. This core formation stage is predicted to be followed by a process called the gravothermal catastrophe, or core-collapse, which produces a small dense core in the center of the halo \cite{kochanek2000,balberg2002}. The dynamic evolution predicted by SIDM models may resolve several contradictions between cold dark matter (CDM) models and observational data, such as the diversity problem (CDM; \cite{KuziodeNaray_2014,oman15,errani18,read19,Relatores_2019,santos20,Hayashi_2020,Li_2020,cruz_2025}, SIDM; \cite{kaplinghat16,kamada17,sameie20,correa21,carton2022,correa22,Nadler_2023,dYang_2023, carton_simpaper, carton2024, roberts2024}) and the cuspy halo problem (CDM; \cite{flores94,moore94,NFW97,moore99,weinberg15,10.1111/j.1365-2966.2004.07836.x}, SIDM; \cite{Kaplinghat_2020,Bhattacharyya_2022}).

One potential probe of SIDM is substructure lensing, where we observe strong gravitational lenses and infer properties of the substructure inside the lens and along the line-of-sight \cite{mao1998,Dalal_2002,nierenberg_2014,Hezaveh_2016a,Hezaveh_2016b,nierenberg_2017,daylan2018,gilman_2019,gilman_lens2021,dhanasingham_2022,dhanasingham_2023,vegetti2023stronggravitationallensingprobe,Wagner-Carena_2023,gilman2025jwstlensedquasardark}. These tests are highly sensitive to the core-collapsed dark matter halos predicted by many SIDM models, which would strongly perturb the lensing potential of their host \cite{vegetti2023stronggravitationallensingprobe,kong2025stronglensingperturberssidm,gannon2025darkmattersubstructurelensing}. Observations of strong gravitational lenses have found evidence for subhalo perturbers, through detection of isolated line-of-sight structure \cite{vegetti_2010a,vegetti_2010b,vegetti_2012,vegetti_2014,nierenberg_2014,Hezaveh_2016a,Li_2025,vegetti_2026} and also measurements of flux ratio anomalies of multiple-image lensed quasars \cite{gilman_2019,pyHalo,gilman2019_masscrel,hsueh_2020,nadler_2021,gilman_2022,gilman2025jwstlensedquasardark}.

Inferring the properties of dark substructure from gravitational lenses requires accurate modeling of both the lens substructure and the line-of-sight structure for a given model of dark matter. Recent work has demonstrated that line-of-sight structure imparts a unique signature on the lensing potential. This signature can be understood through ``effective multiplane gravitational lensing,'' a method in which we define an effective gravitational lensing potential which incorporates the cumulative effects of line-of-sight structure on the lensing deflection map \cite{dhanasingham_2022,dhanasingham_2023,dhanasingham2025neuralposteriorestimationlineofsight}. The statistic of interest in this analysis is the effective lensing potential's two-point correlation function, which is projected against Chebyshev polynomials and quantified by the relative strength of each multipole moment.

While these observational probes are promising tests of dark matter structure, the constraints they provide rely on the robustness of our substructure and line-of-sight structure models. For SIDM in particular, accurate modeling of the rapid and drastic gravothermal catastrophe is crucial to accurate predictions for lensing observables. Current approaches to core-collapse models involve a number of choices and simplifying assumptions regarding the collapsing halo profile, such as assuming the core-collapsing halo instantaneously transitions from its initial density profile to the final core-collapsed state. As an additional challenge, the timescale on which a given halo collapses is often dominated by uncertainty. Some of this uncertainty is from a lack of numerical precision \cite{mace2024,Palubski_2024}, and some stems from the complex evolution of the systems we are trying to model \cite{gannon2025darkmattersubstructurelensing,carton2024,kamionkowski2025numericalevolutionselfgravitatinghalos}. It is not immediately clear what the best modeling choices are, and what effect various simplifying assumptions have on the complicated pipeline of substructure lensing inference. This uncertainty is a disconnect between existing inference and the underlying physical models they aim to test. Without a solid understanding of this problem and guidance for clear modeling choices, the ability of lensing observations to illuminate the nature of dark matter will be limited.

To quantify this modeling challenge, we test various modifications of the SIDM core-collapsing halo density profile in the context of substructure lensing. These modifications represent modeling choices that can be made when generating a subhalo and line-of-sight halo population for a substructure lensing inference study. Our modifications include variation of both the initial halo (uncollapsed) and final (core-collapsed) density profiles, and comparisons between instantaneously collapsing and time-evolving halo profiles. Our baseline core-collapse model is a gravothermal fluid solution calibrated to $N$-body simulations of isolated SIDM halos, and model tidal evolution based on SIDM substructure simulations. We implement our model variations into random realizations of main lens subhalos and line-of-sight structure generated with the open source package \texttt{pyHalo} \cite{pyHalo}, assuming a Planck 2018 cosmology \cite{planck_2018}. We vary the density profiles of perturbing substructure, testing their impact on effective multiplane gravitational lensing predictions. 

This paper is structured as follow. In Section \ref{sec:methods} we detail how we generate realizations for lensed systems, implement various SIDM models, and perform gravitational lensing calculations. Next we present our calibration to $N$-body substructure simulations (Section \ref{sec:nbody_results}), and show results for our density profile variations (Sections \ref{sec:CCvar_results} and \ref{sec:binary_results}). We summarize our findings in Section \ref{sec:conclusion}.

\section{Methods}

\label{sec:methods}
In this section we describe how we generate realizations of galaxy-scale lens systems with subhalos and line-of-sight halos, model core-collapse in these realizations, and produce lensing statistics from these models. First, we include details of the $N$-body simulations we use to model tidal effects (Section \ref{sec:nbody}), and how we generate the CDM subhalos and line-of-sight halos for our lensing realizations (Section \ref{sec:pyhalo}). Next, we replace the CDM halos with SIDM halos by modifying the density profiles. We discuss our various SIDM profiles, starting with our continuously collapsing model (Section \ref{sec:shengqi}) and variations to the collapse threshold (Section \ref{sec:CCtimevar}), and moving on to models with binary core-collapse (Section \ref{sec:binary}). We end the section with an overview of effective multi-plane gravitational lensing, which we use to calculate the two-point correlation function of the lensing convergence (Section \ref{sec:lensing}).

\subsection{N-body Simulations}
\label{sec:nbody}

To validate our core-collapse model (described in detail in Section \ref{sec:shengqi}), we use $N$-body simulation data first generated as part of Ref. \cite{carton_simpaper}. These data were generated with the $N$-body code \texttt{Arepo} \cite{arepo2010}, using a previously tested SIDM module \cite{vogelsberger2012,vogelsberger13,vogelsberger14}.

The simulations feature an analytic host profile and four different random realizations of substructure generated with \texttt{Galacticus}\footnote{\url{https://github.com/galacticusorg/galacticus}}, a semi-analytic model of galaxy formation \cite{galacticus}. The analytic host is modeled as an NFW profile with mass $M_{200c}=10^{13}\ M_\odot$ and concentration $c_{200}=6.7$ at a redshift of $z_\mathrm{lens}=0.5$, and the merger trees include subhalos and sub-subhalos with infall masses above $10^8\ M_\odot$. We define the virial mass $M_{200c}$ as the mass of a spherical overdensity defined such that the halo density is equal to 200 times the critical density of the Universe. To accurately model low mass subhalos that may be insufficiently resolved for accurate core-collapse behavior, these simulations employ a hierarchical simulation framework. In this approach we group subhalos by halo mass and simulate them in batches of differing mass resolution, ensuring that subhalos are similarly resolved on all scales. More details on the simulation methods are available in Ref. \cite{carton_simpaper}.

Each set of initial conditions is run in \texttt{Arepo} with a variety of SIDM models. In this work, we are using simulations with SIDM scattering cross-sections given by:

\begin{equation}
    \sigma/m = \frac{\sigma_0}{\left(1+v_\mathrm{rel}^2/\omega^2\right)^2}
    \label{eqn:vdep_sigmaM}
\end{equation}
where $v_\mathrm{rel}$ is the relative velocity between two dark matter particles, and the characteristic cross-section $\sigma_0$ and cutoff velocity $\omega$ are free model parameters \cite{YangD_2023,gilman_lens2021,slone_2023,outmezguine_2023,shengqi_ccmodel,zhongYM_2023}. This study uses the simulations for only two sets of cross-section model parameters: \{$\sigma_0=200\ \mathrm{cm}^2/\mathrm{g}$, $\omega=50\ \mathrm{km}/\mathrm{s}$\} and \{$\sigma_0=200\ \mathrm{cm}^2/\mathrm{g}$, $\omega=200\ \mathrm{km}/\mathrm{s}$\}. We use these simulations to calibrate an approximate treatment of tidal effects in gravothermal core-collapse, which is discussed in more detail in Section \ref{sec:shengqi}.

\subsection{Gravitational Lens Realizations}
\label{sec:pyhalo}

We generate realizations of gravitational lens subhalos and line-of-sight halos using version 1.4.1 of the Python package \texttt{pyHalo}\footnote{\url{https://github.com/dangilman/pyHalo}}, an open-source toolkit for generating dark matter halo populations for gravitational lensing calculations \cite{pyHalo}. Our realizations are generated with the preset CDM model in \texttt{pyHalo}, which allows for custom settings adjusting the lens system, source system, and population parameters for subhalos and line-of-sight halos. Each of our realizations is for a lens system at redshift $z_\mathrm{lens}=0.5$. The lens host halo has a mass of $M_{200c}=10^{13}\ M_\odot$ and concentration of $c_{200}=6.7$, matching the host halo in the $N$-body simulations (see Section \ref{sec:nbody}). The main deflector is modeled as a power-law ellipsoid with Einstein radius $\theta_{\rm E}=1.0\ \mathrm{arcsec}$, eccentricity components $(e_1=0.05,\ e_2=0.08)$, and logarithmic slope $2.0$ \cite{Auger_2010}. External gravitational shear components are set to $(\gamma_1=0.01,\ \gamma_2=-0.01)$. The source is placed at $z_\mathrm{source}=1.0$, which is a typical source redshift for galaxy-scale strong lenses \cite{wu2022_quasarCat,lyu2025largeskyareamultiobject} and comparable to data included in recent James Webb Space Telescope (\textit{JWST}) strong lens analysis \cite{gilman2025jwstlensedquasardark}.

Subhalos of the main deflector in our $N$-body simulation comparison (Section \ref{sec:nbody_results}) are rendered in the mass range $10^8-10^{10}\ M_\odot$ to match the range of substructure mass in the $N$-body simulations. For the rest of the study (Section \ref{sec:CCvar_results} and onward) we extend to lower masses, and render subhalos and line-of-sight halos in the range $10^6-10^{10}\ M_\odot$. Structure below $10^6\ M_\odot$ has relatively little impact on lensing measurements \cite{gilman2025jwstlensedquasardark}, and structure more massive than $10^{10}\ M_\odot$ can often be modeled individually due to visible baryons or significant lensing effects (see Ref. \cite{birrer_2019}). We sample subhalos from a power-law mass function:
\begin{equation}
    \frac{\mathrm{d}^2 N_\mathrm{sub}}{\mathrm{d}m\ \mathrm{d}A}=\frac{\Sigma_\mathrm{sub}}{m_0}\left(\frac{m}{m_0}\right)^\alpha\mathcal{F}(M_\mathrm{host},z)
    \label{eqn:shmf}
\end{equation}
with logarithmic slope $\alpha=-1.90$ and pivot mass $m_0=10^8\ M_\odot$ based on $N$-body simulations \cite{springelAq_2008,bensonSHMF_2020}. We set the subhalo mass function normalization to $\Sigma_\mathrm{sub}=0.0033\ \mathrm{kpc}^{-2}$, which produces subhalo projected densities similar to the \texttt{Galacticus} merger trees used in our $N$-body comparison (see Section \ref{sec:nbody_results}). This value is about a factor of 30 lower than predictions by $N$-body simulations \cite{NadlerSymphony_2023} and early gravitational lensing inference from \textit{JWST} \cite{gilman2025jwstlensedquasardark}, but we do not expect rescaling of the mass function to affect our primary results regarding the relative impact of density profile features. The final term in Equation \eqref{eqn:shmf} captures scaling with the host halo mass and redshift:
\begin{equation}
    \mathrm{log}_{10}(\mathcal{F})=k_1\mathrm{log}_{10}\left(\frac{M_\mathrm{host}}{10^{13}\ M_\odot}\right) + k_2\mathrm{log}_{10}(z_\mathrm{lens}+0.5)
\end{equation}
with $k_1=0.88$ and $k_2=1.7$ based on calibration to \texttt{Galacticus} \cite{gilman2019_masscrel}.

Following the same procedure as Ref. \cite{dhanasingham_2022}, we render line-of-sight halos in a double-sided cone with a 6 arcsec opening angle except when otherwise specified. Halo populations at each redshift are sampled from a rescaled Sheth-Tormen (ST) mass function \cite{shethTormen_2001,gilman_2019}:

\begin{equation}
    \frac{\mathrm{d}^2N_\mathrm{LOS}}{\mathrm{d}m\mathrm{d}V}=\delta_\mathrm{LOS}(1+\xi_\mathrm{2halo})\left|\frac{\mathrm{d}^2N}{\mathrm{d}m\mathrm{d}V}\right|_\mathrm{ST},
    \label{eqn:ST}
\end{equation}
where, $\delta_\mathrm{LOS}$ is a normalization factor which we fix at 1.0 and $\xi_\mathrm{2halo}$ is the two-halo term of the three-dimensional two-point correlation function. This two-halo correction describes the correlation of structure near the main deflector which is not already captured in the host halo's bound structure. The final term in Equation \eqref{eqn:ST} is the standard Sheth-Tormen mass function.

We generate formation redshifts for subhalos and line-of-sight halos using the method described in Appendix C of Ref. \cite{fvdbHaloForm_2014}, which models the mass accretion and potential well growth histories of dark matter halos. Infall redshifts for subhalos are sampled from the distribution in the \texttt{Galacticus} merger trees used in our $N$-body simulations. This is a mass independent distribution of redshifts, calculated numerically and implemented into \texttt{pyHalo}. The mass added by substructure and line-of-sight structure is offset by sheets of negative lensing convergence at various redshifts, added by \texttt{pyHalo}. We consider only subhalos with infall times greater than $4\ \mathrm{Gyr}$, to match the population used in Ref. \cite{carton_simpaper} as precisely as possible.

In the CDM halo realization, subhalo and line-of-sight halo density profiles are modeled as a truncated Navarro-Frenk-White (NFW) profile \cite{Baltz_2009}:

\begin{equation}
    \rho_\mathrm{TNFW}(r)=\frac{\rho_\mathrm{s}}{r/r_\mathrm{s}(1+r/r_\mathrm{s})^2}\frac{r_\mathrm{t}^2}{r^2+r_\mathrm{t}^2}
    \label{eqn:tNFW}
\end{equation}
where $\rho_\mathrm{s}$ and $r_\mathrm{s}$ are the NFW scale density and radius respectively \cite{NFW96}. The truncation radius $r_\mathrm{t}$ for subhalos is determined by the tidal track predictions in Ref. \cite{xiaolong_galtrunc}, which are matched to \texttt{Galacticus} predictions (using the \texttt{TruncationGalacticus} class in \texttt{pyHalo}). Line-of-sight halos are truncated at $r_{\rm t}=r_{50}$ (where the enclosed density is 50 times the critical density of the Universe at the halo's redshift), which is comparable to the splashback radius \cite{more2015_splashback} (using the \texttt{TruncationRN} class in \texttt{pyHalo}).

\subsection{Continuous Core-collapse}
\label{sec:shengqi}

\subsubsection{Density Profile}

We implement smoothly evolving SIDM halo density profiles using the empirical model presented in Ref. \cite{shengqi_ccmodel}, with the addition of a truncation radius $r_{\rm t}$ (see Appendix \ref{sec:truncated_profile} for more detail on the truncated profile). This density evolution is piecewise in time, depending on whether the halo is in the early stages of evolution (core-formation) or late stages (core-collapse):

\begin{align}
\nonumber
\hat{\rho}_\mathrm{early}(\hat{r},\beta\hat{\sigma}\hat{t}) &=
     \frac{\mathrm{tanh}(\hat{r}/r_\mathrm{c})}{\hat{r}(1+\hat{r})^2}\frac{\hat{r}_\mathrm{t}^2}{\hat{r}^2+\hat{r}_\mathrm{t}^2} \\
    \hat{\rho}_\mathrm{late}(\hat{r},\beta\hat{\sigma}\hat{t})&=\frac{\rho_\mathrm{c}}{1+(\hat{r}/r_\mathrm{c})^{2.19}(1+\hat{r}/r_\mathrm{out})^{0.81}}\frac{\hat{r}_\mathrm{t}^2}{\hat{r}^2+\hat{r}_\mathrm{t}^2}.
    \label{eqn:shengqiCC}
\end{align}
In this paper, hatted variables (such as $\hat{\rho}$ and $\hat{r}$) are dimensionless and are calculated by rescaling by the initial NFW profile parameters. Whether we are using $\hat{\rho}_\mathrm{early}$ or $\hat{\rho}_\mathrm{late}$ depends on how far the halo is into its core-collapse evolution, parametrized by the product $\beta\hat{\sigma}\hat{t}$. We calculate this timescale from the dimensionless scattering cross-section $\hat{\sigma}=r_{\rm s}\rho_{\rm s}\sigma/m$ (a rescaling of the SIDM interaction cross-section $\sigma/m)$, the dimensionless time parameter $\hat{t}=t\sqrt{4\pi G\rho_{\rm s}}$ (where $t$ is the age of the halo), and the heat transfer calibration parameter $\beta$ (a dimensionless parameter calibrated to $N$-body simulations \cite{mace2025calibratingsidmgravothermalcatastrophe}). The parameters $\rho_\mathrm{c}$, $r_\mathrm{c}$, and $r_\mathrm{out}$ are all dimensionless functions of the rescaled evolution time $\beta\hat{\sigma}\hat{t}$, with full expressions given in the appendix of Ref. \cite{shengqi_ccmodel}. In this work we neglect the effect of the substructure density profile on the truncation, and choose the truncation radius $\hat{r}_{\rm t}=r_{\rm t}/r_{\rm s}$ to match the truncation radius used in the CDM realization generated by \texttt{pyHalo}, described in Section \ref{sec:pyhalo}.

The profile evolution switches from early to late evolution at the time when the core is at its largest, $\mathrm{log}_{10}(\beta\hat{\sigma}\hat{t})=1.341$. The evolution terminates deep into the core-collapse regime at $\mathrm{log}_{10}(\beta\hat{\sigma}\hat{t})=2.237$, after which we freeze the density profile in time. This time corresponds to the final snapshot from our gravothermal fluid code solution, but does not necessarily correspond to the true final state of a core-collapsed halo. We experiment with shifting this ending time in Section \ref{sec:CCtimevar}.

\subsubsection{Deflection Angle Profile}

To make gravitational lensing predictions we must compute the deflection angle profile, which is calculated from the enclosed projected mass within a distance $r$ from the halo center \cite{gilman_lens2021}:

\begin{equation}
    \alpha(r)\propto \frac{1}{r}\int_0^r \mathrm{d}r_\mathrm{2D} \ r_\mathrm{2D}\int_{-\infty}^\infty\mathrm{d}z\ \rho\left(\sqrt{r_\mathrm{2D}^2+z^2}\right).
    \label{eqn:alphaInt}
\end{equation}
We have not found a closed-form solution to this integral for our density profile, so we proceed numerically.

To calculate the deflection angle as a function of radius, $\beta\hat{\sigma}\hat{t}$, and truncation radius, we create a look-up table for Equation \eqref{eqn:shengqiCC}. We use 150 uniformly distributed points over $\mathrm{log}_{10}(\hat{r})\in[-2,3]$, 271 points over $\mathrm{log}_{10}(\beta\hat{\sigma}\hat{t})\in[-5.6,2.237]$\footnote{Steps in $\mathrm{log}_{10}(\beta\hat{\sigma}\hat{t})$ are not evenly spaced, but are instead based on the variation in the density profile parameters $\mathrm{log}_{10}(\rho_\mathrm{c})$, $\mathrm{log}_{10}(r_\mathrm{c})$, and $\mathrm{log}_{10}(r_\mathrm{out})$. A new step in time is added to the look-up table only if one of these parameters would change by at least 0.5\% of their total range over the full halo evolution.}, and 150 uniformly distributed points over $\mathrm{\log}_{10}(\hat{r}_{\rm t})\in[-2,3]$, and linearly interpolate between points on this three-dimensional grid. We numerically integrate to produce Equation \eqref{eqn:alphaInt} over this grid, creating a look-up table for the deflection angle as a function of $\hat{r}$, $\beta\hat{\sigma}\hat{t}$, and $\hat{r}_{\rm t}$.

\subsubsection{Time evolution}
\label{sec:pyhalo_timeEvol}

To retrieve the deflection angle profile for a particular subhalo or line-of-sight halo, we first determine $\beta\hat{\sigma}\hat{t}$ by calculating the age and halo parameters for that individual halo, placing it at the appropriate position along its unique core-collapse timeline. It is straightforward to calculate $\hat{t}$ for each halo, using the formation redshift and the redshift of observation, which is generated individually for each line-of-sight halo and is $z_\mathrm{lens}$ for all subhalos.

To calculate $\hat{\sigma}$, we bypass the complex velocity dependence in the cross-section (Equation \eqref{eqn:vdep_sigmaM}) by constructing a characteristic velocity that is representative of typical interparticle relative velocities in a given subhalo. We then compute $\hat{\sigma}$ for that one velocity, as an approximation to the subhalo's evolution with a velocity-dependent cross-section. This is a reasonable approximation for a single halo, where the inter-particle velocities will be distributed around such a characteristic velocity. The characteristic velocity is calculated as $\langle v_\mathrm{rel}\rangle=3.8\sigma_v$, where $\sigma_v$ is the one dimensional velocity dispersion of the initial NFW halo at $r=r_{\rm s}$. Integrating over the Boltzmann distribution would suggest a characteristic velocity of $2.26\sigma_v$, but $N$-body simulations \cite{carton_simpaper} and gravothermal fluid modeling \cite{shengqi_ccmodel,outmezguine_2023} have shown that a factor of $3.8$ better models the effective heat capacity of the dark matter halo. Once $\langle v_\mathrm{rel}\rangle$ is calculated for a given halo, the characteristic $\sigma/m$ is found using Equation \eqref{eqn:vdep_sigmaM}. With $\sigma/m$ we can place the subhalo on the evolutionary track in Equation \eqref{eqn:shengqiCC}, and determine the appropriate density profile.

The empirical model in this subsection is valid for isolated dark matter halos, and does not include environmental effects that could accelerate or delay the gravothermal evolution. We hypothesize that the effect of a subhalo's environment on the core-collapse rate can be approximated with a numerical factor altering the dimensionless time parameter in the empirical model, and that this factor can be measured from $N$-body simulations of subhalo evolution.

To account for tidal effects, we break the time evolution of each subhalo into two phases: one before infall and one after. Before infall, each halo evolves according to our isolated model. The post-infall evolution will require calibration to the $N$-body simulations discussed in Section \ref{sec:nbody}.

\begin{figure}
\begin{subfigure}[t]{\columnwidth}
    \centering
	\includegraphics[width=\textwidth]{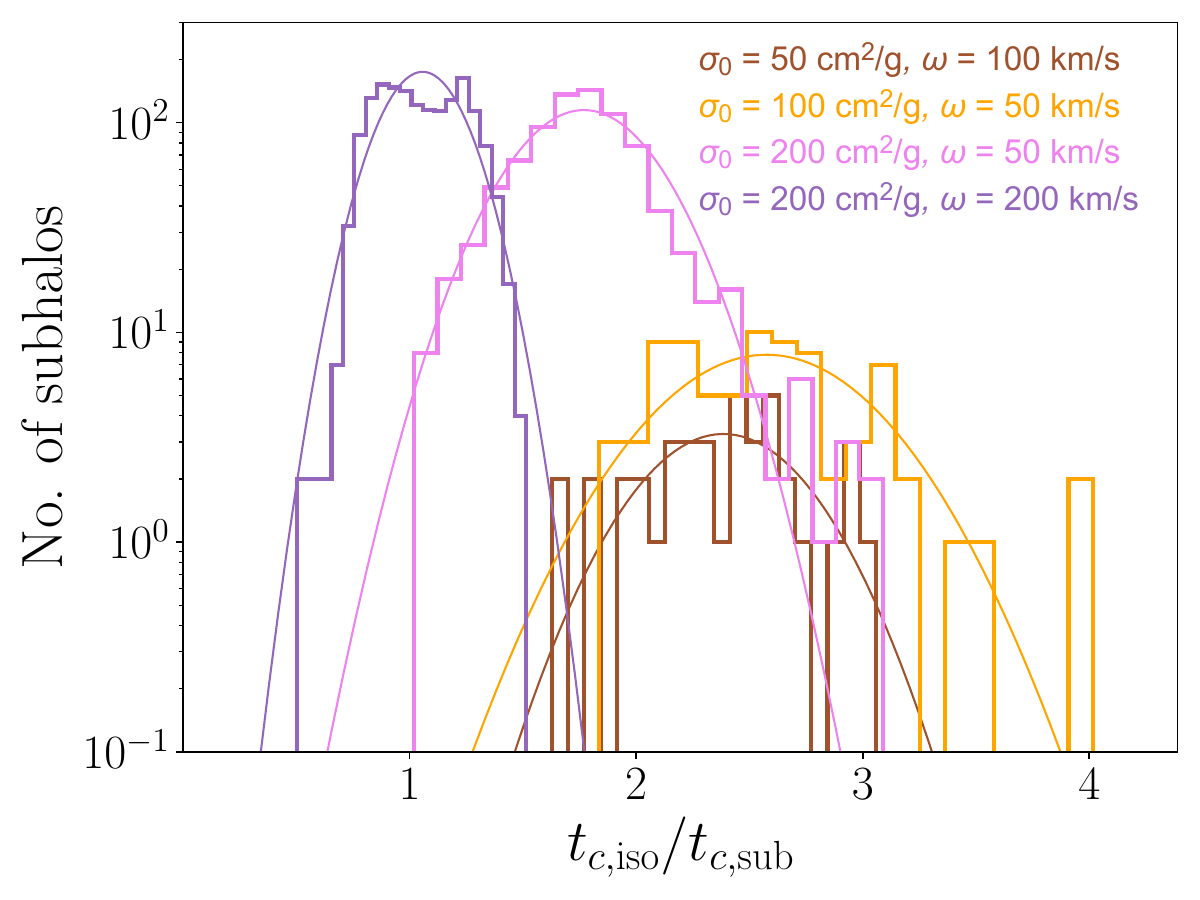}
\end{subfigure}
    \caption{The distribution of core-collapse acceleration factors from N-body simulations, generated using data from Ref. \cite{carton_simpaper}. In this work we study only the \{$\sigma_0=200\ \mathrm{cm}^2/\mathrm{g}$, $\omega=50\ \mathrm{km}/\mathrm{s}$\} and \{$\sigma_0=200\ \mathrm{cm}^2/\mathrm{g}$, $\omega=200\ \mathrm{km}/\mathrm{s}$\} models, which have relatively fast core-collapse and a large number of collapsed halos to sample the collapse distribution. The lines are Gaussian fits, with best fit mean and standard deviation values given in Table \ref{tab:gauss_fits}.}   \label{fig:tidalAcc_dists}
\end{figure}

\begin{center}
\begin{table}
\begin{tabular}{|c |c || c | c|} 
 \hline
 $\sigma_0$ ($\mathrm{cm}^2/\mathrm{g}$) & $\omega$ ($\mathrm{km}/\mathrm{s}$) & Mean & Standard Deviation \\
 \hline\hline
  50 & 100 & 2.39 & 0.359 \\ 
 \hline
 100 & 50 & 2.58 & 0.439 \\ 
 \hline
 200 & 50 & 1.77 & 0.302 \\ 
 \hline
 200 & 200 & 1.06 & 0.185 \\ 
 \hline
\end{tabular}
\caption{Means and standard deviations for the Gaussian fits to the $t_\mathrm{c,iso}/t_\mathrm{c,sub}$ distributions shown in Figure \ref{fig:tidalAcc_dists}.}
\label{tab:gauss_fits}
\end{table}
\end{center}

We use the $N$-body simulations from Ref.~\cite{carton_simpaper} to sample collapse acceleration (or deceleration) factors for each halo, defined as the ratio between the the isolated halo's collapse time ($t_\mathrm{c,iso}$) and the collapse time of that same halo evolved in a host potential ($t_\mathrm{c,sub}$). We expect these measurements to be somewhat biased towards halos that collapse faster, since we exclude any halos which do not collapse in isolation. Figure \ref{fig:tidalAcc_dists} shows those distributions for four velocity-dependent SIDM models, including the two we select for this study. We find that the acceleration factors for both selected models are well modeled by Gaussian distributions, with fit parameters listed in Table \ref{tab:gauss_fits}. For each subhalo in our \texttt{pyHalo} realizations, we sample an acceleration factor from the Gaussian distribution and apply it to the post-infall phase of the SIDM evolution. In addition to the influence of the tidal field, we expect this acceleration factor to include the effects of highly energetic collisions between subhalo and host particles, which can lead to evaporation \cite{carton_simpaper,carton2024,Nadler_2020}. While we take a probabilistic approach here and do not track individual orbits, orbital history plays a major role in the tidal evolution of subhalos. A more precise calculation of tidal acceleration would consider the orbital parameters of each subhalo. This more sophisticated approach is beyond the scope of this study, but may be possible with a better understanding of SIDM tidal tracks \cite{carton2024}.

\subsubsection{Collapse Threshold Variation}
\label{sec:CCtimevar}

\begin{figure}
\begin{subfigure}[t]{\columnwidth}
    \centering
	\includegraphics[trim=25 0 0 0, width=0.95\textwidth]{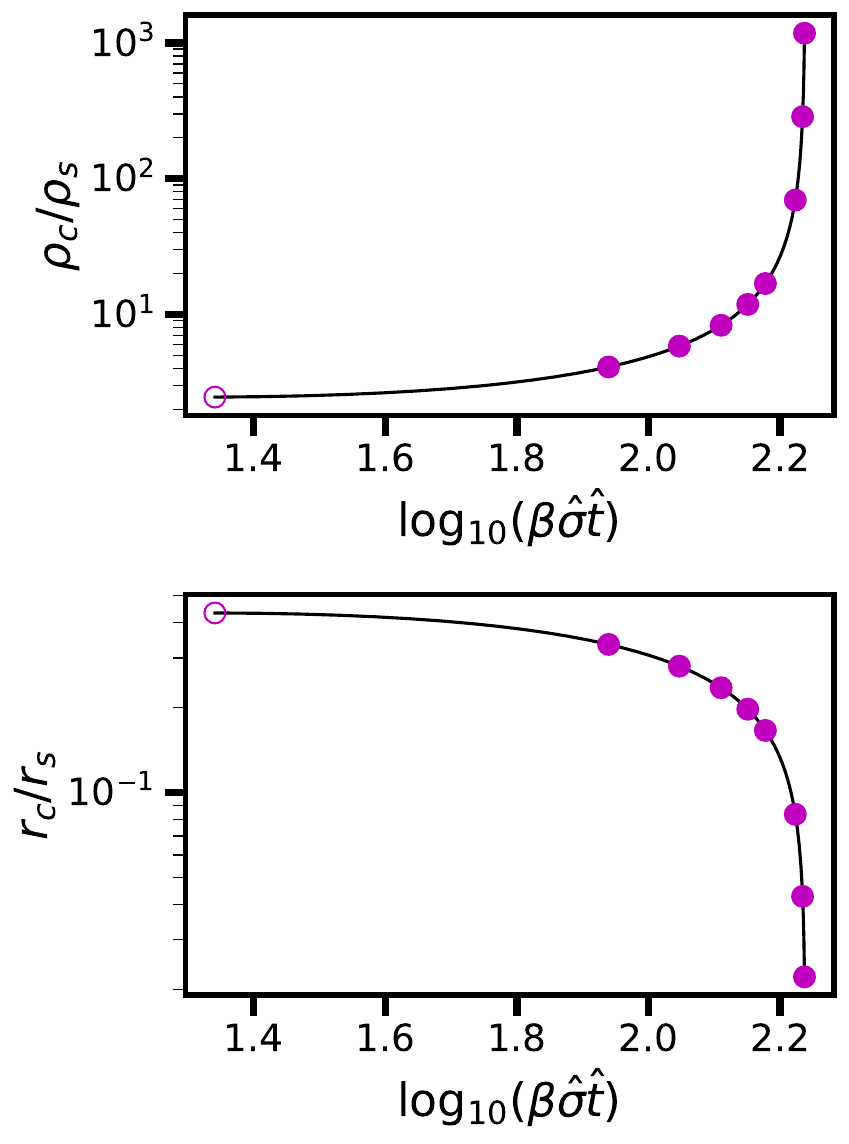}
\end{subfigure}
    \caption{Our selection of core-collapse thresholds as discussed in Section \ref{sec:binary}. The top panel shows the halo central density ($\rho_{\rm c}$) evolution, the bottom shows the core radius ($r_{\rm c}$), and $\mathrm{\log}(\beta\hat{\sigma}\hat{t})$ is the dimensionless evolution time (see Equation \eqref{eqn:shengqiCC}). The circles indicate the points in time at which we sample the density profile for binary collapse. The empty circle is at the maximal core time, and the filled circles are our eight choices for collapse times in Table \ref{tab:finalSnaps}.}   \label{fig:initFinal}
\end{figure}

While our gravothermal solution ends with a central density that diverges as the collapse time is approached, the true final state of a core-collapsed halo remains unknown. Core-collapse may be halted at some point as the gravothermal evolution is interrupted by other physical processes, such as black hole formation \cite{balberg2002,Feng_2022,roberts_2025,Jiang_2026}. To test how the lensing signal responds to different core-collapse end states, we pick a range of final points to consider. Table \ref{tab:finalSnaps} lists our eight different choices of ending times, along with their values of $\rho_\mathrm{c}$ and $r_\mathrm{c}$ (Equation \eqref{eqn:shengqiCC}), which are displayed in Figure \ref{fig:initFinal}. %We pick these five points such that $\log_{10}(\rho_c)$ changes by $20\%$ between each.

For each termination point, we can model collapsed halos in two ways. The first approach is to \emph{freeze} the density profiles once they reach the final time, using the density profile at the termination time as the fully collapsed profile. When varying the collapse time in this way, we are testing the effects of halting core-collapse earlier. We expect this approach to reduce the lensing power, since we are lowering the central density of all collapsed halos. The second approach is to \emph{boost} the core-collapsed profile up to the fully collapsed $\mathrm{log}(\beta\hat{\sigma}\hat{t})=2.237$ profile once they pass the termination time. This method keeps the final density profile the same regardless of our choice for the collapse time threshold, but allows nearly collapsed halos to be treated as collapsed instead of suppressing them back down to the initial profile. Unlike the freezing approach, where we suppress lensing power, this approach would enhance it.
\begin{center}
\begin{table}
\begin{tabular}{|c | c | c|| c| c|} 
 \hline
 $\mathrm{log}_{10}(\beta\hat{\sigma}\hat{t})$ & $\rho_\mathrm{c}/\rho_\mathrm{s}$ & $r_\mathrm{c}/r_\mathrm{s}$ & $\mathrm{log}(\rho_\mathrm{c})/\mathrm{log}(\rho_\mathrm{f})$ & $t_\mathrm{c}/t_\mathrm{f}$ \\
 \hline\hline
 1.939 & 4.114 & 0.3350 & 20\% & 50.4\% \\ 
 \hline
 2.047 & 5.858 & 0.2805 & 25\% & 64.5\% \\ 
 \hline
 2.110 & 8.343 & 0.2351 & 30\% & 74.7\% \\ 
 \hline
 2.151 & 11.88 & 0.1974 & 35\% & 82.0\% \\ 
 \hline
 2.178 & 16.92 & 0.1659 & 40\% & 87.3\% \\ 
 \hline
 2.223 & 69.61 & 0.08364 & 60\% & 96.8\% \\
 \hline
 2.234 & 286.3 & 0.04277 & 80\% & 99.3\%  \\
 \hline
 2.237 & 1178 & 0.0222 & 100\% & 100\% \\ 
 \hline
\end{tabular}
\caption{Final density profile choices for collapse threshold variation. The model parameters for each choice are listed on the left side of the table, and the values of the central density ($\rho_\mathrm{c}$) and time ($t_\mathrm{c}$) compared to the full evolution model ($\rho_\mathrm{f}$ and $t_\mathrm{f}$) are on the right.}
\label{tab:finalSnaps}
\end{table}
\end{center}
\subsection{Binary Core-collapse}
\label{sec:binary}
A common model of SIDM core-collapse in substructure lensing is to treat the core-collapse process as instantaneous, and model each halo as one of two profiles: one for collapsed halos, and another for un-collapsed halos \cite{gilman_lens2021,gilman_resCC2023,gilman2025jwstlensedquasardark}. If the two density profiles have analytically solvable deflection angle integrals (Equation \eqref{eqn:alphaInt}) this treatment can reduce computational costs. Binary collapse models assume the core formation stage of SIDM evolution does not significantly alter lensing measurements on a population level. To test this assumption and gauge the sensitivity of gravitational lensing to the intermediate stages of SIDM halo evolution, we implement several models of binary core-collapse and compare them to our continuously evolving model (Section \ref{sec:shengqi}).

To model binary collapse in our gravitational lenses, we select the initial and final profiles from our continuously evolving gravothermal profile (Section \ref{sec:shengqi}). With the most basic implementation of this method, any halo with $\mathrm{log}(\beta\hat{\sigma}\hat{t})<2.237$ is modeled with the initial TNFW profile, while halos with $\mathrm{log}(\beta\hat{\sigma}\hat{t})\geq2.237$ are considered collapsed, and modeled with Equation \eqref{eqn:shengqiCC} at $\mathrm{log}(\beta\hat{\sigma}\hat{t})=2.237$.

In some of our binary collapse tests, we randomly select which halos core-collapse, instead of tracking each halo's evolution individually. This probabilistic approach has been used with binary collapse models in the literature \cite{gilman_lens2021,gilman_resCC2023}. To determine the collapse probabilities, we bin subhalos by infall mass and calculate the collapse fraction in each bin using the individually evolved halo sample. We then randomly select halos in each mass bin for core-collapse, using the desired collapse fractions as the collapse probability. Collapse probabilities for line-of-sight halos are calculated similarly, but are binned separately from subhalos due to the expected influence of the host halo on subhalo collapse rates \cite{nishikawa_2020,correa21,carton2022,carton_simpaper}. For both subhalos and line-of-sight halos, we use the mass bins $[10^6-10^7\ M_\odot]$, $[10^7-10^8\ M_\odot]$, $[10^8-10^9\ M_\odot]$, and $[10^9-10^{10}\ M_\odot]$. Unlike the most recent probabilistic collapse models in the literature \cite{gilman_resCC2023}, this method of random collapse does not account for any individual halo properties, such as concentration and halo formation time. By using this simple model for core-collapse we intend to set an upper bound on inaccuracies caused by collapse randomization, which we expect more sophisticated collapse choices would improve upon.

We also experiment with different initial profiles. The standard approach is to set an un-collapsed dark matter halo to the TNFW initial condition, but for some cross-sections and halo populations there may be many halos that reach the cored phase without core-collapsing. To test the best initial profile for binary collapse, we use the profile of maximum core size ($\mathrm{log}_{10}(\beta\hat{\sigma}\hat{t})=1.341$) for un-collapsed halos in some of our tests. The $\rho_{\rm c}$ and $r_{\rm c}$ of the cored initial profile are shown as empty circles in Figure \ref{fig:initFinal}.

\begin{figure*}
        \centering
        \includegraphics[height=4.9in]{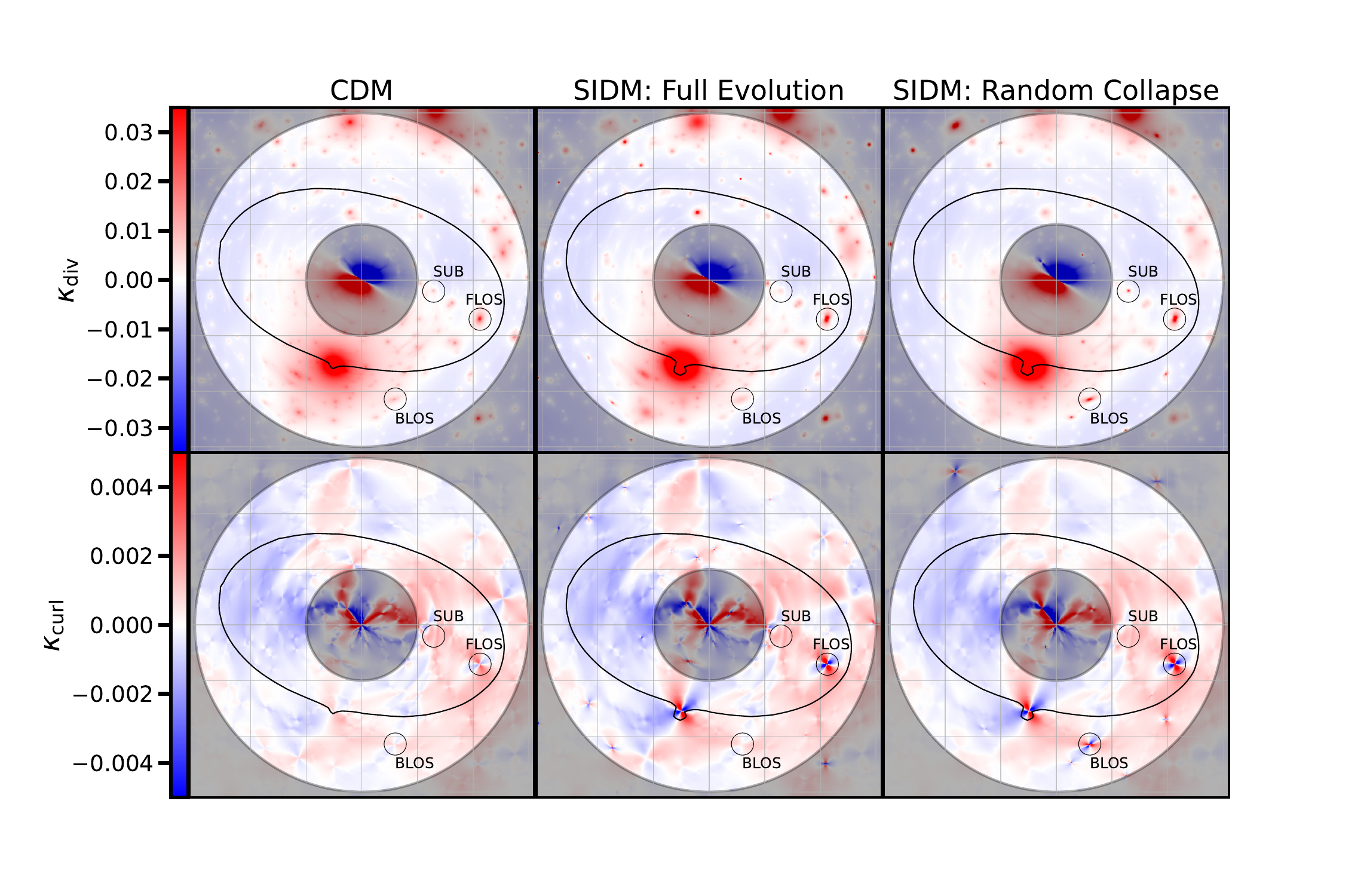}

    \caption{$\kappa_\mathrm{div}$ (top) and $\kappa_\mathrm{curl}$ (bottom) maps for a single \texttt{pyHalo} rendering, with CDM (left) and two SIDM models (center, right). To isolate the effect of substructure and line-of-sight structure, the convergence of the main lens halo is subtracted out. The SIDM models use a velocity dependent cross-section (Equation \ref{eqn:vdep_sigmaM}) with \{$\sigma_0=200\ \mathrm{cm}^2/\mathrm{g}$, $\omega=200\ \mathrm{km}/\mathrm{s}$\}. Each square of the overlaid grid is $0.5\times0.5\ \mathrm{arcsec}$ in angular size, and shaded regions are excluded by our annular mask when calculating the two-point correlation function. The critical lensing curve is shown in black. Examples of a subhalo (SUB), a background line-of-sight halo (BLOS), and a foreground line-of-sight halo (FLOS) are circled.}\label{fig:convmap}
\end{figure*}

\subsection{Gravitational Lensing}
\label{sec:lensing}

\subsubsection{Effective Multi-plane Lensing}

To quantify the collective contributions of line-of-sight structure to the lensing deflection, we use the effective multi-plane lensing framework detailed in \cite{dhanasingham_2022}. This approach includes the influence of line-of-sight structure through an effective deflection field $\boldsymbol{\alpha}_\mathrm{eff}$:

\begin{equation}
    \boldsymbol{\alpha}_\mathrm{eff}(\textbf{x})=\nabla\phi_\mathrm{eff}(\textbf{x}) + \nabla\times\textbf{A}_\mathrm{eff}(\textbf{x}).
\end{equation}
Unlike in a single-plane lens, which is fully characterized by the scalar potential $\phi_\mathrm{eff}$, the non-linear coupling between lensing planes has given rise to a vector potential $\textbf{A}_\mathrm{eff}$. The observed image can be computed from $\boldsymbol{\alpha}_\mathrm{eff}$ with the lens equation:

\begin{equation}
    \textbf{u}(\textbf{x})=\textbf{x} - \boldsymbol{\alpha}_\mathrm{eff}(\textbf{x})
\end{equation}
where $\textbf{u}$ and $\textbf{x}$ are positions on the source and image plane respectively.

When studying single-plane lensing, it is common to calculate the lensing convergence from the deflection angle $\boldsymbol{\alpha}$ as $\kappa\equiv\frac{1}{2}\nabla\cdot\boldsymbol{\alpha}$. For a single lens plane, the mapping between $\kappa$ and the deflection angle is straightforwardly linked to the gravitational potential \cite{meylan_2006}, and $\boldsymbol{\alpha}$ is explicitly curl-free. In a multi-plane lens the deflection angle now can be sourced by a vector field $\textbf{A}_\mathrm{eff}$ instead of just the gradient of a scalar field, allowing us to define a new convergence based on the curl of the effective deflection field. Our two convergences are calculated as:
\begin{align}
    \kappa_\mathrm{div}&=\frac{1}{2}\nabla\cdot\boldsymbol{\alpha}_\mathrm{eff}-\kappa_0 \\
    \kappa_\mathrm{curl}&=\frac{1}{2}\nabla\times\boldsymbol{\alpha}_\mathrm{eff}\cdot \hat{\textbf{z}},
\end{align}
where $\kappa_0$ is the convergence of the lens system's host halo, and $\hat{\textbf{z}}$ is the line-of-sight direction \cite{gilman_2019,sengul_2020,sengul_2022,dhanasingham_2022}. By subtracting $\kappa_0$ we remove the dominant contribution of the main lens, but preserve the non-linear coupling between the main lens and line-of-sight structure, which leads to the anisotropies that we aim to study. Unlike the single-plane case, neither of these terms are equivalent to the projected mass density. By studying both the divergence and curl convergences, we preserve important information about the non-linear multi-plane coupling \cite{dhanasingham_2022,dhanasingham_2023}.

In this study, we compute $\kappa_\mathrm{div}$ and $\kappa_\mathrm{curl}$ for \texttt{pyHalo} realizations (Section \ref{sec:pyhalo}) with version 1.12.0 of the Python gravitational lensing package \texttt{lenstronomy} \cite{lenstronomy1,lenstronomy2}. We compute the convergence of our $N$-body simulations (Section \ref{sec:nbody}) with the C++ library \texttt{GLAMER} \cite{metcalf_glamer,petkova_glamer}.

\begin{figure}
\begin{subfigure}[t]{\columnwidth}
    \centering
	\includegraphics[width=\textwidth]{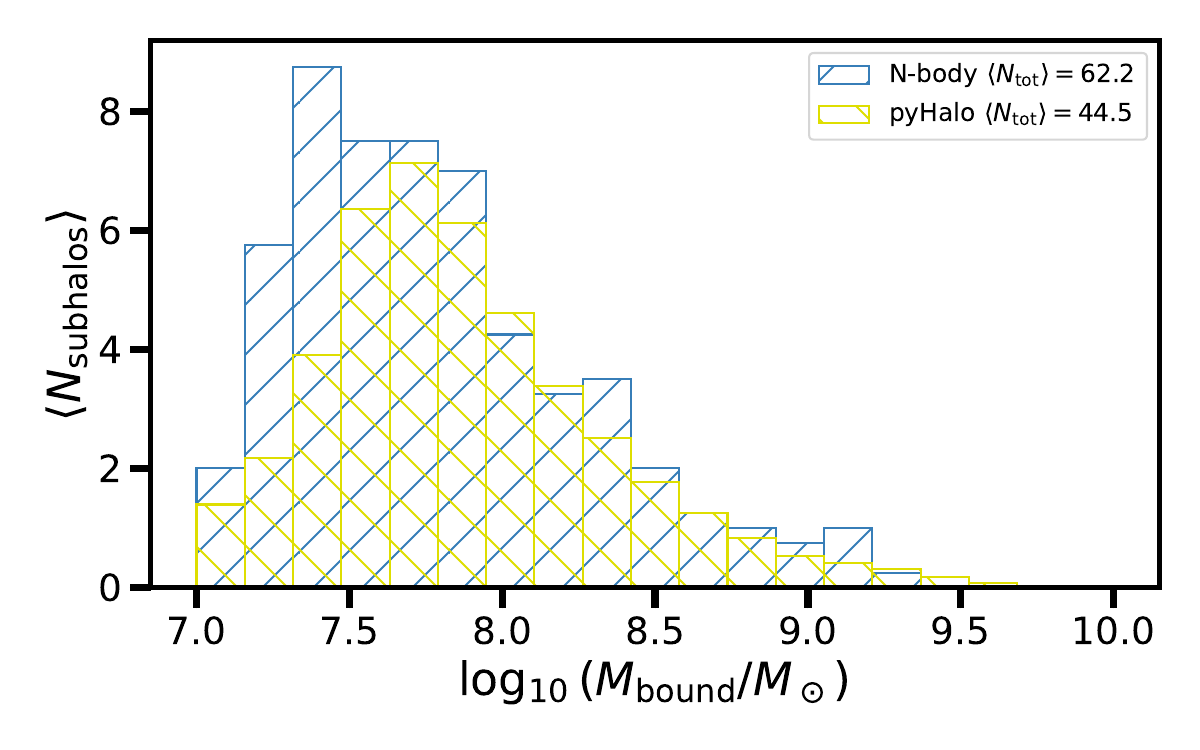}
\end{subfigure}
    \caption{The abundance of subhalos within the 1-10 arcsecond annular mask for our $N$-body simulations and \texttt{pyHalo} realizations. The averages are computed over 4 realizations for our $N$-body simulations (blue), and 200 realizations for our \texttt{pyHalo} data (yellow). The average number $\langle N_\mathrm{tot}\rangle$ of subhalos in the mask for each dataset is given in the legend.}   \label{fig:massfunc}
\end{figure}

\begin{figure*}
    \centering
    \begin{subfigure}{0.5\textwidth}
        \centering
        \includegraphics[height=2.2in]{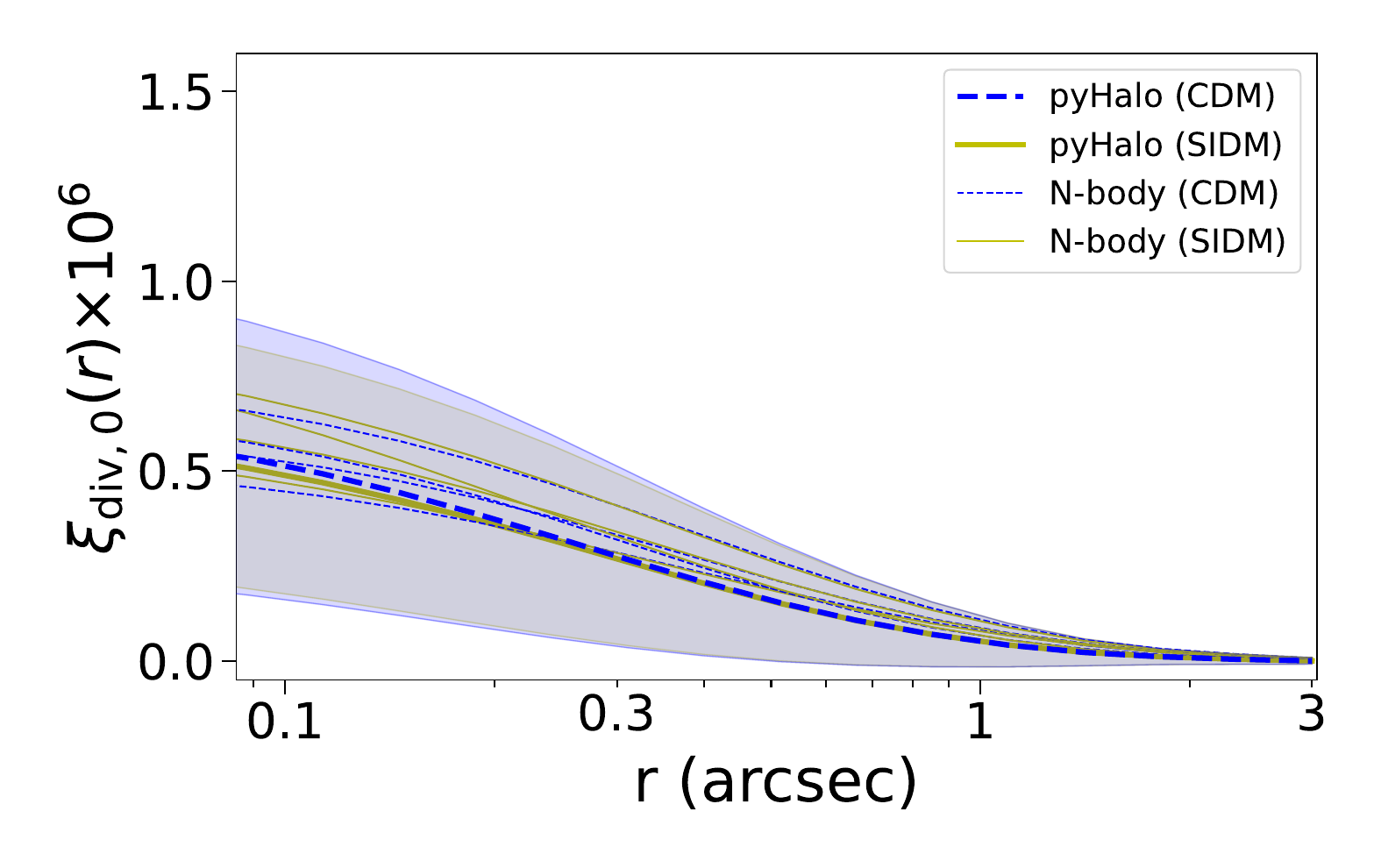}
        \caption{$\sigma_0=200\ \mathrm{cm}^2/\mathrm{g}$, $\omega=50\ \mathrm{km}/\mathrm{s}$}
    \end{subfigure}%
    ~ 
    \begin{subfigure}{0.5\textwidth}
        \centering
        \includegraphics[height=2.2in]{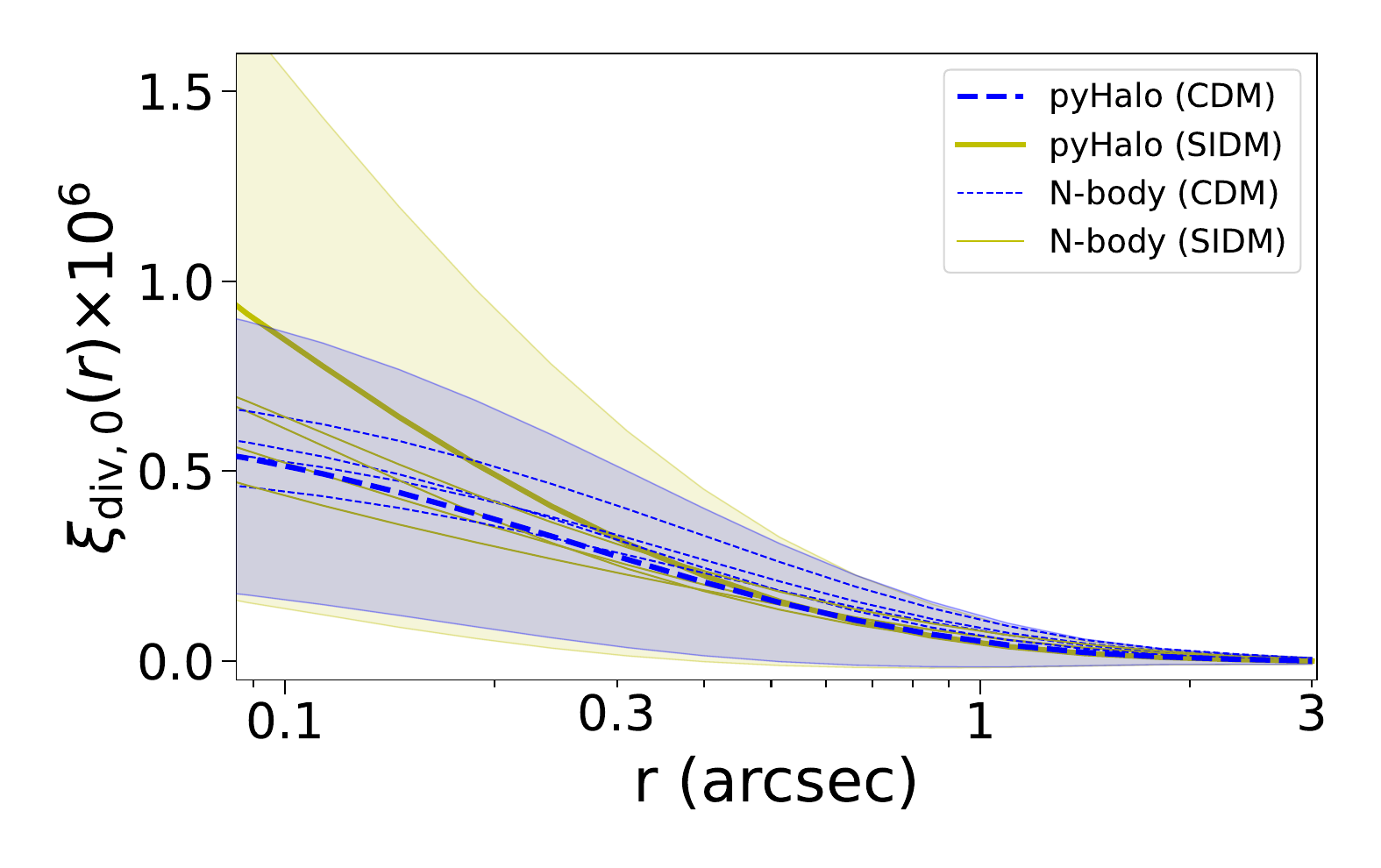}
        \caption{$\sigma_0=200\ \mathrm{cm}^2/\mathrm{g}$, $\omega=200\ \mathrm{km}/\mathrm{s}$}
    \end{subfigure}
    \caption{Comparison of the monopole moment between \texttt{pyHalo} and $N$-body realizations. For the \texttt{pyHalo} realizations, we render 200 random realizations of lens substructure, and show the mean (thick lines) and standard deviation (shaded regions) of $\xi_{\mathrm{div},0}(r)$. For the $N$-body simulations we plot the four individual simulations as thin lines, each averaged over 64 random angles of projection. CDM and SIDM results are distinguished by color and line style, for both the \texttt{pyHalo} renderings and the $N$-body results.}
    \label{fig:nbody_moments}
\end{figure*}

\subsubsection{The Two-point Correlation Function}

The two-point correlation function is a powerful summary statistic that encodes information about the distribution of substructure, as well as their inner deflection angle profiles (and hence their density profiles) \cite{Hezaveh_2016a,bayer_2023,chatterjee_2017,diazRivero_2018a,diazRivero_2018b,brennan_2019,cyrRacine_2019,sengul_2020,dhanasingham_2022,dhanasingham_2023}. In this section we will compute the correlation function for an effective multiplane lens, and use it to disentangle line-of-sight effects from the influence of the main lensing plane. As shown in \cite{dhanasingham_2023}, the line-of-sight halos create structures in the $\kappa_\mathrm{div}$ field that exhibit even parity under the transformation $\theta\rightarrow-\theta$, but the corresponding structures in $\kappa_\mathrm{curl}$ exhibit odd parity. To measure these signals we will compute the two-point correlation functions of $\kappa_\mathrm{div}$ and $\kappa_\mathrm{curl}$, and decompose them onto orthonormal bases. With the parity of line-of-sight signatures in mind, we will decompose $\kappa_\mathrm{div}$ onto an even basis and $\kappa_\mathrm{curl}$ onto an odd basis.

For a convergence map $\kappa$, the two-point correlation function as a function of a vector $\boldsymbol{r}$ linking two points in the image plane is:
\begin{equation}
    \xi(\boldsymbol{r})=\frac{1}{S}\int_S\mathrm{d}^2\boldsymbol{r}_1[\kappa(\boldsymbol{r}_1)-\langle\kappa(\boldsymbol{r}_1)\rangle][\kappa(\boldsymbol{r}_2)-\langle\kappa(\boldsymbol{r}_2)\rangle]
\end{equation}
where $\boldsymbol{r}_2=\boldsymbol{r}_1+\boldsymbol{r}$ and $S$ is the area over which we are calculating the correlation function. For our density profile modeling tests, we compute $\xi(\boldsymbol{r})$ over an annular mask with an inner radius of $0.5\ \mathrm{arcsec}$ and outer radius $1.5\ \mathrm{arcsec}$, chosen to focus on structure near the lensing critical curve. In this study we are primarily interested in the how the shape of the two-point correlation function changes under variations in perturbing halo density profiles, but the mass sheet transform would only rescale the amplitude by a constant factor. We are thus ignoring the mass sheet degeneracy, as in Ref. \cite{dhanasingham_2023}.

We decompose $\xi_\mathrm{div}(r,\theta)$ onto an orthonormal basis of even functions $\cos(\ell\theta)$ as:
\begin{equation}
    \xi_\mathrm{div}(r,\theta)=\sum_{\ell=0}^\infty \xi_{\mathrm{div},\ell}(r)\cos(\ell\theta),
\end{equation}
with the correlation multipoles given by:
\begin{equation}
\label{eqn:div_moments}
    \xi_{\mathrm{div},\ell}(r)=\frac{2-\delta_{\ell0}}{\pi}\int_0^\pi \mathrm{d}\theta\ \xi_\mathrm{div}(r,\theta)\cos(\ell\theta),
\end{equation}
and $\delta_{ij}$ denoting the Kronecker delta.

Due to the odd parity of the line-of-sight structure in $\kappa_\mathrm{curl}$, we will decompose $\xi_\mathrm{curl}(r,\theta)$ onto an orthogonal basis of odd functions $B_\theta\sin(\ell\theta)$ with:
\begin{equation}
    B_\theta=
    \begin{cases}
        +1 &  0 \leq \theta < \pi/2 \\
        -1 & \pi/2 \leq \theta < \pi.
    \end{cases}
\end{equation}
This decomposition gives:
\begin{equation}
    \xi_\mathrm{curl}(r,\theta)=\sum_{\ell=1}^\infty\xi_{\mathrm{curl},\ell}(r)B_\theta\sin(\ell\theta),
\end{equation}
with the correlation multipoles given by:
\begin{equation}
\label{eqn:curl_moments}
    \xi_{\mathrm{curl},\ell}(r)=\frac{2}{\pi}\int_0^\pi \mathrm{d}\theta\ \xi_\mathrm{curl}(r,\theta)B_\theta\sin(\ell\theta).
\end{equation}
We will use the correlation multipoles in Equations \eqref{eqn:div_moments} and \eqref{eqn:curl_moments} as our primary comparison of lensing statistics between realizations. In this study we will be concerned with the monopole and quadrupole moments of the $\kappa_\mathrm{div}$ map ($\xi_{\mathrm{div},0}$ and $\xi_{\mathrm{div},2}$), and the quadrupole moment of the $\kappa_\mathrm{curl}$ map ($\xi_{\mathrm{curl},2}$).

Figure \ref{fig:convmap} shows an example of the $\kappa_\mathrm{div}$ and $\kappa_\mathrm{curl}$ maps for three different models: CDM, SIDM with full halo evolution (Section \ref{sec:shengqi}), and SIDM with randomly selected binary collapse (Section \ref{sec:binary}). We indicate a main deflector subhalo, a foreground line-of-sight halo (between the observer and the main deflector), and a background line-of-sight halo (between the source and the main deflector) on the convergence maps as examples of the distinct structure contributed by each. In this figure we can see the quadrupolar structures in the $\kappa_\mathrm{div}$ field from line-of-sight halos, which are intensified for core-collapsed halos. This quadrupole contribution is captured in the $\xi_{\mathrm{div},2}$ measurement, which we will show in the next section. Additionally, the signal from line-of-sight halos are often stretched into arcs due to the non-linear coupling between lens planes.

When comparing the CDM convergence maps (left column) in Figure \ref{fig:convmap} to the SIDM full evolution (center panels), we can see the signals of many halos changing. Some halos core-collapse, such as our example foreground line-of-sight halo, producing high-density peaks in the convergence maps that induce relatively strong lensing effects. Some, like the indicated subhalo and background line-of-sight halo, happen to be in the cored phase of SIDM evolution and have weaker signatures compared to the cuspy CDM profiles. In the random binary collapse case (right column) the collapse state of each halo is randomized, with some producing similar signals as in the full evolution, some becoming collapsed, and some becoming cored.

\section{Results}

\label{sec:results}

\subsection{Comparison to N-body Simulations}
\label{sec:nbody_results}
To validate our core-collapse acceleration model we generate 200 \texttt{pyHalo} realizations with the same subhalo infall mass range as the $N$-body simulations ($\mathrm{log}_{10}(M/M_\odot)\in[8,10]$), and compare the subhalo population and monopole moment of the convergence map to that of the $N$-body simulations. We do not include line-of-sight structure in these simulations, since the $N$-body simulations produce halos only in the main lens plane. We compute maps of the effective lensing convergence for the \texttt{pyHalo} and $N$-body lenses with \texttt{lenstronomy} and \texttt{GLAMER} respectively, over $800\times800$ pixel, $20\times20\ \mathrm{arcsec}$ grids. We then use \texttt{pyHalo} to compute compute correlation multipoles from the effective convergence maps, using an annular aperture with inner radius $1\ \mathrm{arcsec}$ and outer radius $10\ \mathrm{arcsec}$. This aperture is significantly larger than the one we will use in the next section, because the $N$-body simulations include only subhalos with infall masses above $10^8\ \mathrm{M}_\odot$. Using a larger aperture provides enough subhalos to make meaningful measurements of the lensing statistics.

By varying the subhalo mass function normalization $\Sigma_\mathrm{sub}$ (see Section \ref{sec:pyhalo} for the definition), we can match the \texttt{pyHalo} lens realizations to the subhalo population at the end of the CDM N-body simulation. With the choice of $\Sigma_\mathrm{sub}=0.0033$ our \texttt{pyHalo} lenses broadly reproduce the subhalo population of our $N$-body simulations, as shown in Figure \ref{fig:massfunc}. We are unable to produce the population of very low mass subhalos seen in the $N$-body simulations however, with our realizations producing relatively few $M_\mathrm{bound}<10^{7.5}M_\odot$ subhalos and about 18 fewer subhalos in the annular region on average. These very low mass subhalos are likely the most severely tidally stripped subhalos, suggesting the treatment of tidal effects in \texttt{pyHalo} differs from the CDM simulation results.

\begin{figure*}
    \centering
	\includegraphics[width=\textwidth]{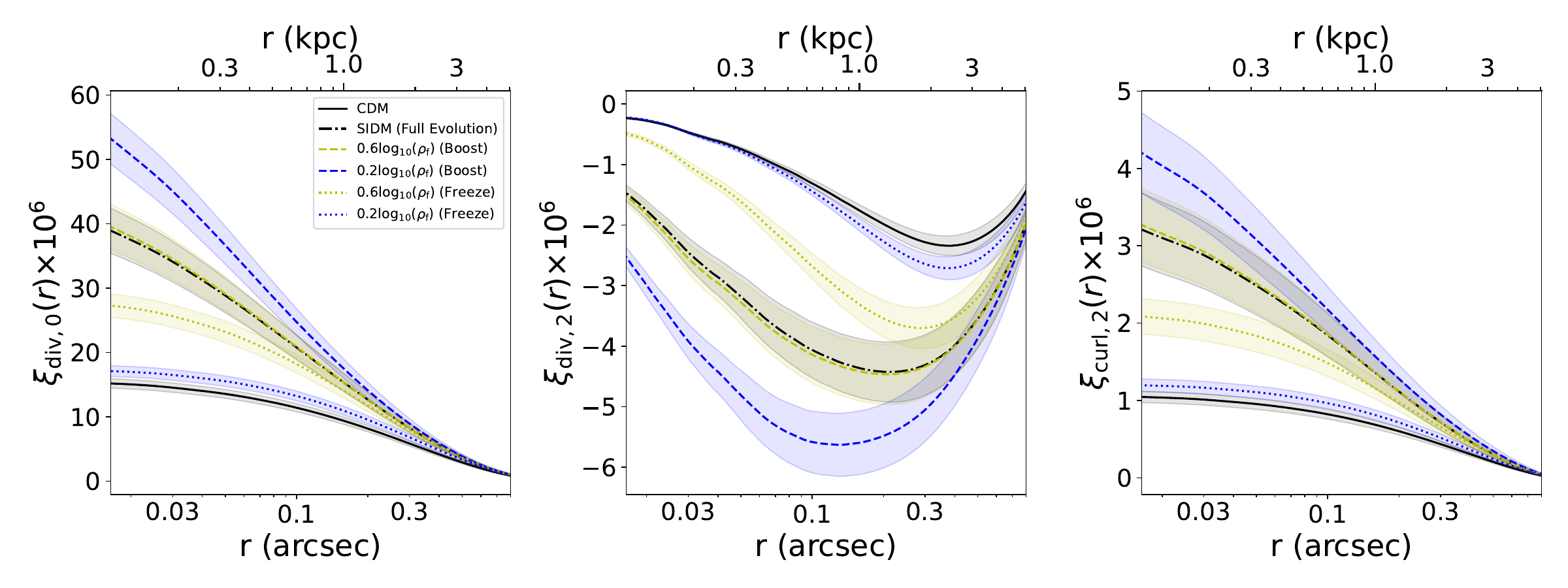}
    \caption{Multipole moments for CDM, SIDM full evolution, and SIDM early collapse models with two collapse thresholds from Table \ref{tab:finalSnaps}. The SIDM models use a velocity dependent cross-section (Equation \ref{eqn:vdep_sigmaM}) with \{$\sigma_0=200\ \mathrm{cm}^2/\mathrm{g}$, $\omega=200\ \mathrm{km}/\mathrm{s}$\}. Lines are the mean multipole moments as a function of distance, and shaded regions are the uncertainty of the mean derived from 200 realizations. Note that the distance axis is in angular units on the bottom of the figure, and in physical units on the top.}   \label{fig:fvb_tall}
\end{figure*}

Figure \ref{fig:nbody_moments} shows the monopole moment results for two different SIDM models. Since we only have four samples of $N$-body lenses, we consider whether or not they could reasonably be considered as samples from the set of \texttt{pyHalo} renderings as a test of our tidal acceleration approach.
The $N$-body simulation monopole moments are consistent with the \texttt{pyHalo} results in all cases, with the $N$-body moments falling within one standard deviation of the \texttt{pyHalo} mean moments. We note, however, that the SIDM results with the \{$\sigma_0=200\ \mathrm{cm}^2/\mathrm{g}$, $\omega=50\ \mathrm{km}/\mathrm{s}$\} model are not distinguishable from the CDM results in both \texttt{pyHalo} and the $N$-body simulations. This is consistent with the results of Ref. \cite{carton_simpaper}, which found that core-collapse in this model is uncommon at halo masses above $10^9 \ M_\odot$. For the rest of our analysis we will focus on the \{$\sigma_0=200\ \mathrm{cm}^2/\mathrm{g}$ $\omega=200\ \mathrm{km}/\mathrm{s}$\} model, which produces more significant deviations from CDM to test as we modify the core-collapse evolution.

While the $N$-body simulations are consistent with \texttt{pyHalo} realizations, this does not confirm that our approximate treatment of tidal effects is sufficient for all applications. The $N$-body simulations represent only four samples of the full distribution of possible subhalo populations. Based on Figure \ref{fig:nbody_moments}, we find that our tidal acceleration treatment is a reasonable first approximation to the true tidal behavior of core-collapsing subhalos, and we can reasonably proceed with it in this study. In Section \ref{sec:conclusion} we discuss how our tidal acceleration model affects the results of this study, and how subhalos could be modeled more accurately.

\subsection{Collapse Time Variation}
\label{sec:CCvar_results}
\begin{figure}
\begin{subfigure}[t]{\columnwidth}
    \centering
	\includegraphics[trim=25 0 0 10, width=0.95\textwidth]{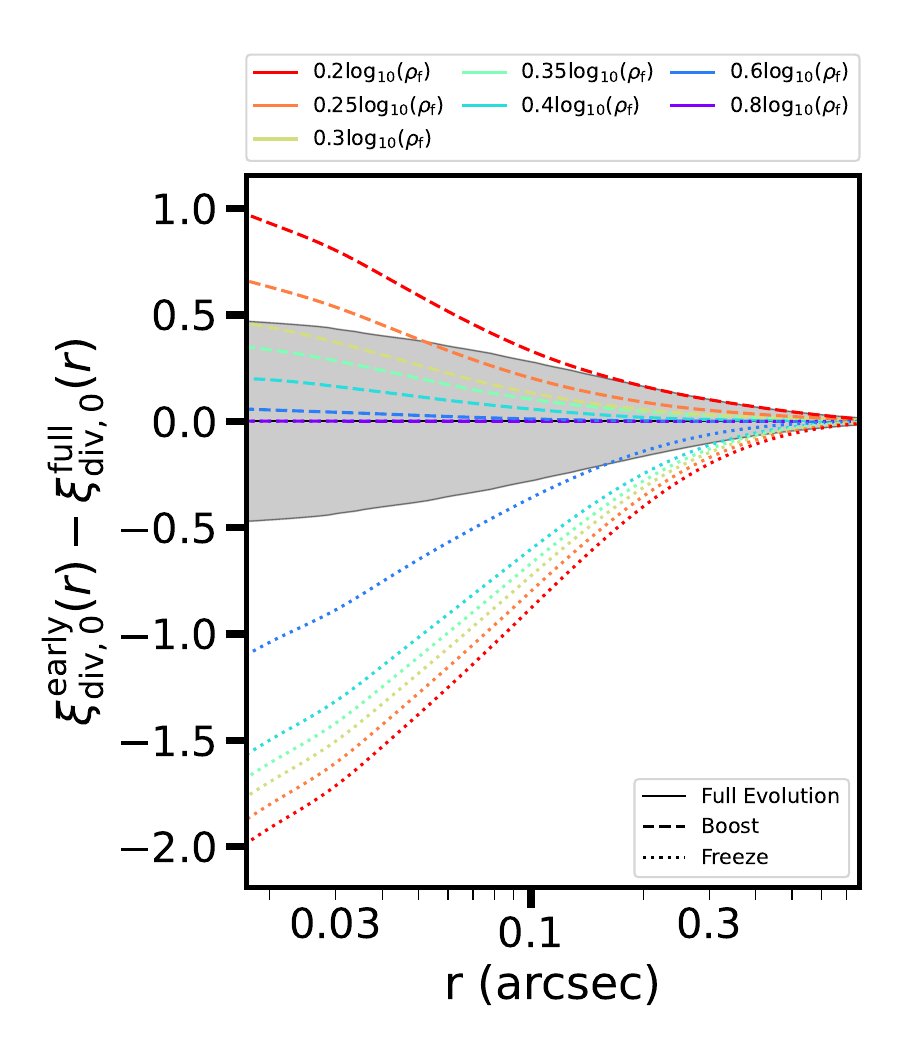}
\end{subfigure}
    \caption{Relative difference between the average monopole moment of full SIDM evolution ($\xi^\mathrm{full}_{\mathrm{div},0}$) and early collapse ($\xi^\mathrm{early}_{\mathrm{div},0}$). The lines are the mean values of the monopole moments, with dashed lines boosting the density profile and dotted lines freezing it. The shaded region is the uncertainty in the mean of $\xi^\mathrm{full}_{\mathrm{div},0}$ with 200 realizations.}   \label{fig:fvb_zoom}
\end{figure}

Now that we have checked our core-collapse implementation against the $N$-body data, we test many variations of the SIDM density profile. In this and all following subsections, we calculate lensing statistics over an annular aperture with inner radius $0.5\ \mathrm{arcsec}$ and outer radius $1.5\ \mathrm{arcsec}$, and reduce our lowest rendered halo mass from $10^8\ M_\odot$ to $10^6\ M_\odot$. All of these tests are conducted over the same 200 random realizations of subhalos and line-of-sight halo populations rendered with \texttt{pyHalo}, with convergence maps computed over $300\times 300$ pixel$^2$ ($3.2\times 3.2\ \mathrm{arcsec}^2$) grids. Further details of our substructure and line-of-sight realizations are in Section \ref{sec:pyhalo}. 

In our first test, we vary the collapse time of the continuously evolving density profile. We test two different methods: in one, we freeze the profile at the collapse time, and in the other, we instead boost the profile up to the default final profile ($\mathrm{log}(\beta\hat{\sigma}\hat{t})=2.237$). See Section \ref{sec:shengqi} for details of the continuous profile and these two modifications.

Figure \ref{fig:fvb_tall} shows all three multipole moments for a subset of collapse times in Table \ref{tab:finalSnaps}. In all three subplots we see that for the 60\% collapse threshold, the lensing prediction is significantly different from CDM only in the frozen case, where we halt the density profile evolution once it reaches the threshold density. With the 20\% collapse threshold, both boosting and freezing alter the predictions. Freezing the profile always suppresses the SIDM enhancement, reducing the gap between CDM and SIDM predictions. The boosted density profile, when it has an effect, enhances all three moments. In Figure \ref{fig:fvb_zoom} we show how the monopole moment changes with collapse time, this time plotting every collapse time model. These results follow the general trend found in Figure \ref{fig:fvb_tall}.

From the profile freezing results, we see that altering the central density of core-collapsed halos affects the lensing signal at small scales, for thresholds of $60\%\ \mathrm{log}_{10}(\rho_\mathrm{f})$ and lower. For lower core-collapsed densities, the change becomes significant at smaller scales. The moment $\xi_\mathrm{div,2}(r)$ remains sensitive to density changes at larger scales than the moments $\xi_\mathrm{div,0}(r)$ and $\xi_\mathrm{curl,2}(r)$ do, with distinguishable offsets from full evolution persisting above $0.1\ \mathrm{arcsec}$.

In contrast, the boosting results show that the collapse time has very little effect on the lensing statistics. Since the boosted profile does not affect the monopole moment until our most extreme case ($20\%\ \mathrm{log}_{10}(\rho_\mathrm{f})$), it appears that the final density has a much stronger effect on lensing population statistics than the collapse time.

\begin{figure*}
    \centering
	\includegraphics[width=\textwidth]{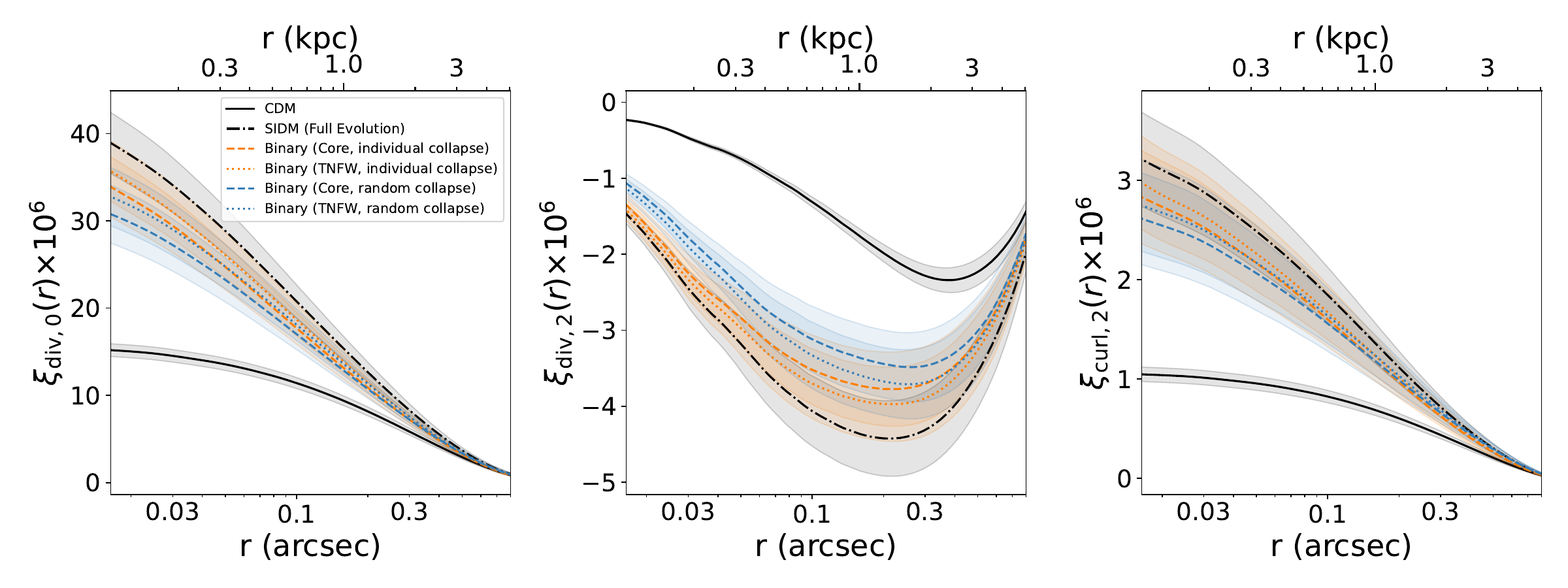}
    \caption{Multipole moments for CDM, SIDM full evolution, and SIDM binary collapse with variations on the collapse method and un-collapsed profile. Among the binary collapse cases, different un-collapsed profiles (cored versus TNFW) are distinguished with line style, and different collapse methods (individual evolution versus random) are distinguished with color. The SIDM models use a velocity dependent cross-section (Equation \ref{eqn:vdep_sigmaM}) with \{$\sigma_0=200\ \mathrm{cm}^2/\mathrm{g}$, $\omega=200\ \mathrm{km}/\mathrm{s}$\}. Lines are the mean multipole moments as a function of distance, and shaded regions are the uncertainty of the mean derived from 200 realizations. Note that the distance axis is in angular units on the bottom of the figure, and in physical units on the top.}   \label{fig:binary_tests}
\end{figure*}

\subsection{Binary Core-collapse}
\label{sec:binary_results}

In our next set of tests, we compare the fully time-dependent density profile (Section \ref{sec:shengqi}) against models with binary core-collapse (Section \ref{sec:binary}). These tests are designed to probe the importance of individual halo modeling when computing lensing population statistics. We compute lensing statistics for CDM, continuous core-collapse, and four variations of binary core-collapse. In the four binary collapse models we vary the un-collapsed density profile (cored versus NFW) and the method of assigning core-collapse (individual evolution times versus randomly assigned collapse). When modeling un-collapsed profiles as cored, we use the maximal cored profile of our gravothermal model (Equation \ref{eqn:shengqiCC} evaluated at $\mathrm{log}(\beta\hat{\sigma}\hat{t})=1.341$).

Figure \ref{fig:binary_tests} shows the three multipole moments for the CDM, SIDM, and various binary collapse cases. In the $\xi_{\mathrm{div},0}$ panel we can see that the small-scale power is suppressed in binary collapse compared to the full time-evolution in all cases. This is not a surprising result, as this binary approach models all halos with the un-collapsed profile if $\mathrm{log}(\beta\hat{\sigma}\hat{t})<2.237$, even if they are partially collapsed. We can also see in this panel that the moment is further suppressed when using cored initial conditions, and when randomly determining halo core-collapse. Since the total fraction of collapsed halos does not change when randomizing core-collapse, this suggests that substructure with a more pronounced lensing effect tends to collapse faster. Since we do not calculate individual tidal tracks for subhalos, this preferential collapse must be due to halo parameters (mass and concentration) rather than position in the lens.

The center and right panels of Figure \ref{fig:binary_tests} shows the same comparison for the quadrupole moments $\xi_{\mathrm{div},2}$ and $\xi_{\mathrm{curl},2}$ respectively, which probe the line-of-sight structure. The quadrupole moments also show that binary collapse suppresses power on small scales, and this effect is more severe for random collapse and cored initial profiles. Similar to the collapse time tests in Section \ref{sec:CCvar_results}, we see that model differences are distinguishable in $\xi_\mathrm{div,2}(r)$ at larger scales than they are for the other two moments.

All of our binary core-collapse approaches in Figure \ref{fig:binary_tests} suppress the lensing signal on small scales, shifting the multipole moments away from the SIDM full evolution and towards results for CDM. The most accurate binary collapse model, where halo core-collapse is tracked accurately and un-collapsed halos are modeled with a TNFW profile, cannot be clearly distinguished from the full evolution with our sample of 200 strong lenses. The actual detectability of these offsets will depend on the scales being probed in a given analysis, and the magnitude of observational uncertainties.

\section{Summary and Discussion}

\label{sec:conclusion}

In this study, we investigate how the two-point correlation function of the effective lensing convergence field is affected by variations in SIDM core-collapse models. To do this, we have implemented a time-evolving halo density profile fit to gravothermal fluid simulations into the  \texttt{pyHalo} and \texttt{lenstronomy} gravitational lensing codes. To model subhalos with this profile, derived originally for isolated halos, we apply a numerical factor to each halo's core-collapse time based on $N$-body simulations of SIDM subhalos. This factor accounts for the difference in expected collapse time between field halos and substructure, due to tidal effects and evpaoration \cite{carton2022,carton2024}. We have varied the final profile central density and core-collapse time of this continuous model, and modeled the resulting changes to the two-point correlation function. We have also compared the continuous collapse profile to binary collapse models, where each halo is modeled with one of two profiles, depending on whether the halo is considered collapsed or uncollapsed. Our results can be summarized as follows:
\begin{itemize}
    \item The multipole moments of the two-point correlation function are much more sensitive to the density profile of collapsed halos than they are to the collapse timescale (Figure \ref{fig:fvb_zoom}). This suggests modest systematic uncertainties in the SIDM halo core-collapse time (whether the uncertainty is in the definition of the collapse time or in the actual evolution of the density profile) may not significantly affect lensing predictions, as long as the fully collapsed density profile is accurate. With our sample of 200 simulated strong lenses, we find that collapse times up to $~20\%$ shorter than the full evolution do not produce significantly different values of $\xi_{\mathrm{div},0}$ on all scales larger than $2\times10^{-2}\ \mathrm{arcsec}$.
    \item Employing binary core-collapse models suppresses the small-scale enhancement to all multipole moments provided by SIDM core-collapse, but the effect may not be significant at larger scales depending on the number of strong lenses included in the calculation (Figure \ref{fig:binary_tests}). If binary core-collapse models are used, they are most accurate compared to full halo evolution when un-collapsed halos are modeled with TNFW profiles, as opposed to cored profiles which suppress all tested multipole moments.
    \item In this study, random assignment of core-collapsed profiles suppresses the SIDM enhancement of the multipole moments compared to using accurate collapse states for individual halos (Figure \ref{fig:binary_tests}). This suppression suggests a correlation between a halo's influence on the lensing convergence and its core-collapse speed, even if the core-collapse process is ignored. Since we do not model the effect of subhalo position on core-collapse speed, this correlation must be based on the mass and concentration of perturbing halos.
\end{itemize}

The interpretation of our final point requires additional discussion. The collapse acceleration factor we use to approximate tidal effects is randomly determined for each subhalo, and this is likely influencing the results of the random collapse test. Since the subhalos are somewhat randomized in even the full evolution model, we would expect that additional shuffling of collapse states would have a limited effect on the lensing statistics. It is possible that for a fully deterministic model of tidal effects, which considers the orbital parameters of individual halos, random collapse would show a more pronounced difference from the full evolution. This hypothesis could be checked by using more precise SIDM tidal evolution tracks.

The full evolution core-collapse method we present in this manuscript would benefit from further comparisons against common methods in the literature. In Appendix \ref{sec:profiles}, we show that our model is significantly different than the cored and truncated NFW profile, but there are other efficient approximations of core-collapse (such as the cored power-law profile \cite{gilman_lens2021,gilman_resCC2023,dhanasingham_2023} and parametric core-collapse model \cite{Yang_2024}) which could be checked for accuracy against the model in this work, or against other physically motivated full-evolution models. It would also be useful to implement our full evolution model into predictions of flux-ratio anomalies in multiply imaged strong lenses, which can be used to probe substructure density profiles \cite{gilman_lens2021,gilman2025jwstlensedquasardark}. These comparisons can be made in future work, using the framework developed in this study.

Our results show that the two-point correlation function is sensitive to some common simplifications of and uncertainties in the SIDM core-collapse process, with the strongest effects on the smallest distance scales. Any lensing survey that uses this statistic must make intentional and physically motivated choices for substructure and line-of-sight structure halos. It is not yet clear what effect, if any, these subtle changes would have on other measurements such as flux-ratio anomalies which often use a cored and truncated NFW profile (see Appendix \ref{sec:profiles} for additional discussion). To produce confident tests of SIDM, we recommend that these works include analogous testing of their core-collapse profile choices, and compare the resulting deflection angle profiles against physically motivated core-collapse models. Models of SIDM core-collapse in realistic environments must also be refined, as testable predictions of SIDM may have a strong dependence on the final state of core-collapsed halos.

\section*{Acknowledgments}

We thank Daniel Gilman and Tansu Daylan for useful discussions. B. D. and F.-Y. C.-R. acknowledge the support of program HST-AR-17061.001-A whose support was provided by the National Aeronautical and Space Administration (NASA) through a grant from the Space Telescope Science Institute, which is operated by the Association of Universities for Research in Astronomy, Incorporated, under NASA contract NAS5-26555. F.-Y. C.-R. also acknowledges the support of National Science Foundation award OIA-2327192.

C. M. is partially supported by the Presidential Fellowship of the Ohio State University Graduate School. This work benefited from the \emph{Dark Matter Theory, Simulation, and Analysis workshop in the Era of Large Surveys} workshop at the Kavli Institute for Theoretical Physics (NSF PHY-2309135). 

\appendix

\section{Gravothermal Profile Truncation}
\label{sec:truncated_profile}

\begin{figure*}[t!]
        \centering
        \includegraphics[width=\textwidth]{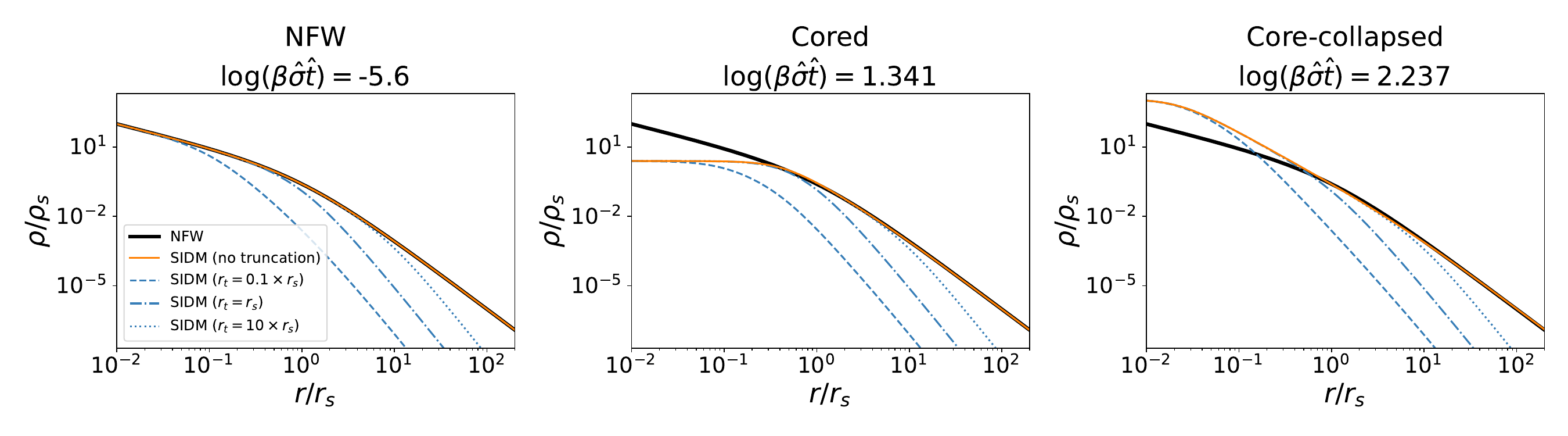}

    \caption{Our truncation method for SIDM core-collapsing halos at three stages of evolution: the initial CDM profile (\emph{left}), the maximally cored profile (\emph{center}), and deeply core-collapsed (\emph{right}). In each panel we compare the gravothermal model density profile without any truncation (orange, Equation \ref{eqn:shengqiCC}) to profiles with three example truncations (blue, truncation differentiated by line style). The initial NFW profile (black) is shown in each panel for reference.}\label{fig:trunc}
\end{figure*}

The gravothermal model fit given in Ref. \cite{shengqi_ccmodel} starts from an NFW profile, and does not include truncation at any time. To model realistic halos with converged total mass, we introduce a truncation to the density profile. We do this by including a common truncation factor used in TNFW profiles $r_{\rm t}^2/(r^2+r_{\rm t}^2)$, with the truncation radius $r_{\rm t}$ as the only additional parameter. We do not explore how this truncation factor may directly affect core-collapse evolution, but account for the effect of tidal interactions on the collapse time by calibrating to $N$-body simulations of subhalo populations (see Section \ref{sec:shengqi} for details). Figure \ref{fig:trunc} shows the truncated density profile for three $r_{\rm t}$ values and three different stages of core-collapse. For actual halos in this study, the truncation radius and core-collapse stage are selected as described in Section \ref{sec:shengqi}.

\section{Density Profile Comparisons}
\label{sec:profiles}

Another choice for binary core-collapse profiles is an NFW-like density profile with truncation and a constant density core (TNFWC)\footnote{This method is implemented in version 1.4.1 of \texttt{pyHalo}, using the TNFWC profile in \texttt{lenstronomy}.}:

\begin{equation}
    \rho_{\mathrm{TNFWC}}(r) = \frac{{\rho^\prime_{\rm s}} {r^\prime_{\rm s}}^3}{\left(r^2+{r^\prime_{\rm c}}^2\right)^{1/2} \left({r^\prime_{\rm s}}^2+r^2\right)} \left(\frac{{r^\prime_{\rm t}}^2}{r^2+{r^\prime_{\rm t}}^2}\right)
    \label{eqn:ctnfw}
\end{equation}
where $\rho^\prime_{\rm s}$ $r^\prime_{\rm s}$, $r^\prime_{\rm c}$, and $r^\prime_{\rm t}$ are the scale density, scale radius, core radius, and truncation radius, respectively. We use primed variables to distinguish from the profile parameters of the original NFW or TNFW profile before core collapse, and from the parameters of a core-collapsed halo with our gravothermal profile.

Both this model and our gravothermal profile (Equation \ref{eqn:shengqiCC}) include a constant density core, but they differ in their outer profiles. The gravothermal profile includes a two-phase outer profile that transitions at the outer radius $r_\mathrm{out}$, while the TNFWC profile does not. This creates a modeling challenge, as it is not clear if there are TNFWC parameter choices that produce similar deflection angles profiles as the gravothermal model. As an exercise, we attempt to match the TNFWC and gravothermal deflection angle profiles with several procedures for selecting TNFWC profile parameters. Our goal is to find TNFWC parameter choices that produce a deflection angle profile similar to a core-collapsed halo with the gravothermal model. We compare the deflection profile instead of the density profile because the deflection angle is the relevant quantity for calculating the convergence map, and includes the effects of 2D projection which are not immediately apparent in a 3D profile. For this test we will use the most collapsed profile in our lookup table ($\log_{10}(\beta\hat{\sigma}\hat{t})=2.237$), and set the truncation radius to be much larger than the NFW scale radius ($r_{\rm t}=r^\prime_{\rm t}=100\times r_{\rm s}$).

The TNFWC profile has four parameters for us to select values for: $\rho^\prime_{\rm s}$ $r^\prime_{\rm s}$, $r^\prime_{\rm c}$, and $r^\prime_{\rm t}$. The scale density $\rho^\prime_{\rm s}$ will be set by mass conservation, where we assume the core-collapse process does not alter the total mass enclosed within some large radius. As we do with our gravothermal profile, we will use $60\times r_{\rm s}$ for this mass conservation radius. We will use a similarly straightforward assumption to set the truncation radius $r^\prime_{\rm t}$, and calculate the truncation radius in the same manner as we do for CDM halos (see Section \ref{sec:shengqi}). Both of these choices are implemented in \texttt{pyHalo}.

The final two parameters are less straightforward. While the un-collapsed TNFW profile and the collapsed TNFWC profile both have a scale radius parameter ($r_{\rm s}$ and $r^\prime_{\rm s}$, respectively), there is no \emph{a priori} reason the collapsed halo will have the same scale length as the un-collapsed profile. One TNFWC core-collapse implementation in \texttt{pyHalo} sets the TNFWC scale radius equal to the core radius ($r^\prime_{\rm s}=r^\prime_{\rm c}$) when the core size is smaller than the truncation radius, and sets it equal to the truncation radius otherwise ($r^\prime_{\rm s}=r^\prime_{\rm t}$). This convention reduces the number of free parameters to one, only the core radius $r^\prime_{\rm c}$, which can then be chosen to represent the desired stage of core-collapse.

\begin{figure*}
    \centering
	\includegraphics[width=\textwidth]{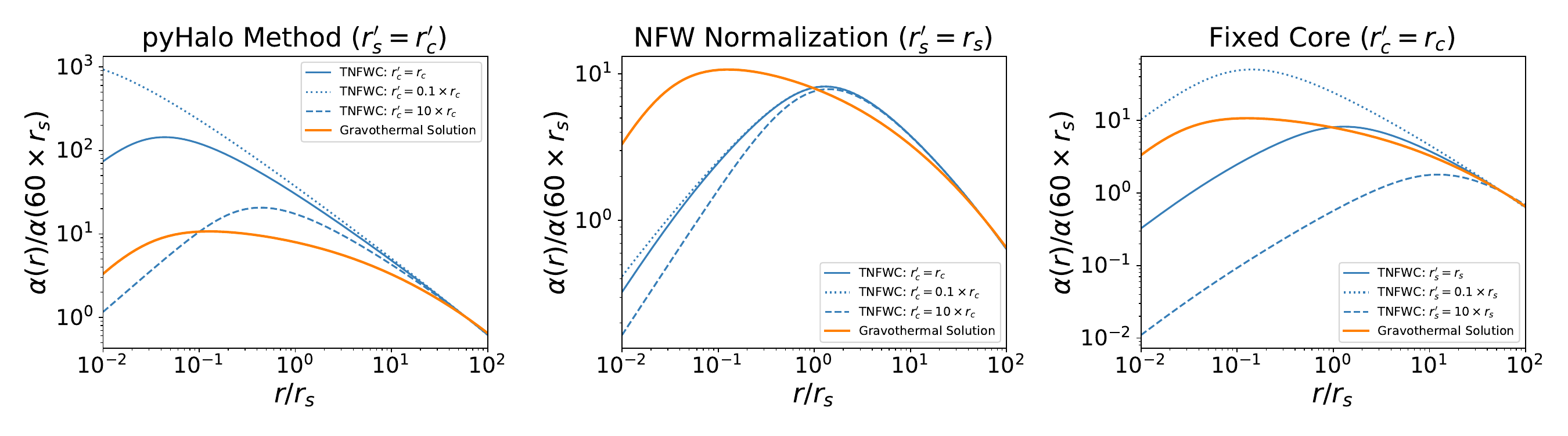}
    \caption{Comparison of deflection angle profiles calculated from the gravothermal solution (Equation \ref{eqn:shengqiCC}) and the TNFWC density profile (Equation \ref{eqn:ctnfw}). All deflection profiles are normalized to be equal at $r=60\times r_{\rm s}$, which is equivalent to setting the enclosed mass enclosed within that radius equal for each halo. The gravothermal deflection profile (bold orange) is for a halo with $\log_{10}(\beta \hat{\sigma} \hat{t})=2.237$, and each panel shows different parameter choices for the TNFWC profile. \emph{Left:} The TNFWC scale radius is set equal to the gravothermal core size, as is done in \texttt{pyHalo} for halos with large truncation radii. The blue lines show the TNFWC deflection profile with three different choices for core size. \emph{Center:} Similar to the left panel, except the TNFWC scale radius is set equal to the NFW scale radius. \emph{Right:} Here we have fixed the TNFWC core radius to the gravothermal core radius, and instead varied the TNFWC scale radius over several values.}
    \label{fig:tnfwc_test}
\end{figure*}

In our first test (left panel of Figure \ref{fig:tnfwc_test}), we adopt \texttt{pyHalo's} convention and set the scale radius equal to the core size ($r^\prime_{\rm s}=r^\prime_{\rm c}$). We then vary the TNFWC core size relative to the gravothermal core size. We see that no choice of $r^\prime_{\rm c}$ produces deflection profiles similar to the gravothermal model; to produce a peak deflection magnitude as low as the gravothermal profile we must use a much larger core, which shifts the peak deflection position to larger radii.

Next, we adopt a new convention and set the TNFWC scale radius equal to the scale length of the original TNFW profile ($r^\prime_{\rm s}=r_{\rm s}$). We vary $r^\prime_{\rm c}$ as before, and again find no satisfactory choice of parameters (center panel of Figure \ref{fig:tnfwc_test}). In this case the core radius has little impact on the deflection profile, with the TNFWC model producing a peak deflection at a much larger radius than the gravothermal model in all cases.

For our final test, we instead fix the core size for the TNFWC profile equal to the gravothermal core size ($r^\prime_{\rm c}=r_{\rm c}$), and instead allow the TNFWC scale radius to vary. This approach comes closest to match the gravothermal profile, with the suggestion that some value $r^\prime_{\rm s}\in[0.1\times r_{\rm s}, \, r_{\rm s}]$ may provide a somewhat close approximation of the gravothermal deflection profile.

Our tests are not exhaustive; we only attempt to match the TNFWC profile to one possible halo, a fully-collapsed halo with a large truncation radius, and more detailed fitting is needed to fully explore the two-dimensional $(r^\prime_{\rm s},r^\prime_{\rm c})$ parameter space. Our findings show that the fitting process is complex, and there is not an immediately clear solution for matching these two profiles. Additionally, we find that using the same TNFWC model assumptions as is done in \texttt{pyHalo} leads to significantly enhanced deflection compared to the gravothermal model's core-collapsed profile. These results do not imply that one profile is more accurate to the core-collapse process than the other, just that the two profiles are not interchangeable. Caution should be exercised when comparing results that do not make the same density profile assumptions.

\bibliography{refs}

\end{document}